# Experimental demonstration of quantum digital signatures using phase-encoded coherent states of light

Patrick J. Clarke[1], Robert J. Collins[1], Vedran Dunjko[1], Erika Andersson[1], John Jeffers[2], Gerald S. Buller[1]

[1]*SUPA, School of Engineering and Physical Sciences, David Brewster Building, Heriot-Watt University, Edinburgh, EH14 4AS, UK.*

[2]*SUPA, Department of Physics, John Anderson Building, University of Strathclyde, 107 Rottenrow, Glasgow, G4 0NG, UK.*

*Correspondence to: r.j.collins@hw.ac.uk.*

**Abstract**: Digital signatures are frequently used in data transfer to prevent impersonation, repudiation and message tampering. Currently used classical digital signature schemes rely on public key encryption techniques, where the complexity of so-called "one-way" mathematical functions is used to provide security over sufficiently long timescales. No mathematical proofs are known for the long-term security of such techniques. Quantum digital signatures offer a means of sending a message which cannot be forged or repudiated, with security verified by information-theoretical limits and quantum mechanics. Here we demonstrate an experimental system which distributes quantum signatures from one sender to two receivers and enables message sending ensured against forging and repudiation. Additionally, we analyse the security of the system in some typical scenarios. The system is based on the interference of phase encoded coherent states of light and our implementation utilises polarisation maintaining optical fibre and photons with a wavelength of 850 nm.

**Introduction**

Alice and Bob have become two of the most important figures in the science of information security, where Alice typically takes the role of sender and Bob receiver. To ensure the validity of important communications, Alice wants to transmit a message to Bob in such a way that he can verify that the message came from her and was not altered in transmission. Additionally it is required that a message that has been validated by one party is further validated by all other parties it is forwarded to. Modern information networks make extensive use of digital signatures to verify the origin and authenticity of messages. These digital signatures are typically based on so-called "trapdoor one-way functions" used in public key cryptography[1-3] which are easy to perform one way but computationally intensive to reverse without prior information. This prior information forms what is known as a "private key", kept secret by Alice and used to decrypt information encrypted using her publicly available "public key". However, there is currently no known proof that reversing such trapdoor one-way functions without the prior information will always be computationally intensive[4] and future advances in mathematical or computer science[5,6] may lead to insecurities in such approaches. Indeed, given enough computational resources the cryptosystem can be broken with current technology[7-9]. In contrast, quantum digital signatures (QDS) offer security verified by information-theoretical limits and quantum mechanics[10,11]. QDS is, roughly speaking, a quantum version of the Lamport public key based scheme for digital signatures[10,12].

As presented here, our experimental setup provides a method for two parties (Bob and Charlie) to receive quantum signatures, which serve as analogues to the public keys in classical cryptography schemes, from an untrusted Alice. These signature states are then used for the full QDS protocol. This allows Alice to sign a message so that it can be validated by Bob and/or Charlie. If an accepted message is forwarded, e.g. from Bob to Charlie, then the forwarded message is guaranteed to also be accepted by Charlie as genuinely coming from Alice. In QDS, security against forging of a message either by Bob or Charlie, or by a fourth external party, is guaranteed as Alice alone has full knowledge about the quantum signature states. Security against repudiation by Alice, in other words that Bob and Charlie will agree on the validity of a forwarded message, is realised, roughly speaking, by ensuring that they have identical quantum signatures. The system must also be robust implying that if all parties act as prescribed by the protocol, classical messages sent from Alice to any single recipient will be confirmed as authentic except with negligible probability in the presence of realistic (experimental) imperfections in equipment. Also, if the authenticated message is forwarded, the message's authenticity will be confirmed except with negligible probability.

In a digital signature scheme, it is vital not only that the public key reveals limited information about the private key, but also that the recipients of the public key can be sure that they have the same public key in order to prevent repudiation of a signature. Classical public keys are readily verified to be identical. Comparing quantum systems, however, is nontrivial[13, 14] and in general difficult to implement[10]. In our system the signatures are encoded as the relative phase shifts of coherent states of light[15]. Quantum comparison of coherent states may be implemented using a 50:50 beamsplitter and has a higher success probability than general comparison methods[16].

Figure 1 (a) shows how two coherent states of light mix on a 50:50 beamsplitter[17]. A coherent state $|\alpha\rangle$ is a quantum state which closely resembles a classical electromagnetic wave. In mathematical terms, $\hat{a}|\alpha\rangle = \alpha|\alpha\rangle$, where $\hat{a}$ is the annihilation operator for the relevant electromagnetic field mode, and $\alpha$ is a complex number. The input states to the beamsplitter are $|\alpha\rangle$ and $|\beta\rangle$ and the output states $|(\alpha-\beta)/\sqrt{2}\rangle$ and $|(\alpha+\beta)/\sqrt{2}\rangle$. Clearly if $\alpha = \beta$ then $|(\alpha-\beta)/\sqrt{2}\rangle = |0\rangle$, the vacuum state, and no light exits through that port. This simple operation forms the basis of the multiport signature comparison system employed by Bob and Charlie, shown in Fig. 1(b). At each receiver the signature from Alice is split into two equal-amplitude components, and one of these is shared with the other receiver, who does the same to their copy of the signature. The retained component of the signature is then mixed on a beamsplitter with the component transmitted from the other receiver. It can be seen from Fig. 1(a) that if the two components are the same, then the original signature will be recovered through one port of the beamsplitter, and $|0\rangle$ will be generated at the other. For any other input state the multiport renders the individual out-bound signature state symmetric with respect to Bob and Charlie. Recall, in order to repudiate a message, Alice has to cause a disagreement between Bob and Charlie concerning the validity of her message. Since the states exiting the multiport are symmetric, with respect to Bob's and Charlie's systems, Bob and Charlie's measurement results will obey the same statistics, and thus if one party validates, so will the other. This is explained in more detail later in this paper.

To utilise the QDS protocol, Alice randomly selects a series of quantum states $|\alpha e^{i\theta}\rangle$ where $\alpha$ is fixed, $\theta \in \{2\pi p/N, p = 0,1,...,N-1\}$, and $N$ is the number of possible phase encodings. The phase of each state is analogous to the classical private key. The principles of quantum mechanics prohibit determining the phases of the states, that is, the private key, with complete certainty if we only have access to the quantum state[18, 14]. Each message bit is signed using a key of length $L$. For a one-bit message $m$, either 0 or 1, Alice generates two sets $\{\rho_{m,k}\}_{k=0}^{L-1}$ of phase-encoded states, with randomly chosen phases defining the corresponding private keys. Alice sends one copy of the pair of sets to Bob and one to Charlie. Bob and Charlie pass the complete series of encoded states $\{\rho_{0,k}\}_k$ and $\{\rho_{1,k}\}_k$ through the multiport of Fig. 1(b). If the original signature states were identical coherent states, this operation will preserve them, otherwise it will symmetrise the overall state shared by Bob and Charlie, which prevents repudiation by Alice. Bob and Charlie then store their phase-encoded states in quantum memory[19]. Quantum memories are a relatively immature technology[23] and have yet to demonstrate long-term storage of quantum states. In our experimental system, Bob and Charlie measure the phase of the laser pulses immediately after they have left the multiport.

In the general case, to send a signed message, Alice sends the message and the classical description of the corresponding private key states to Bob (for example). Bob checks the classical description of the key states against those stored in his memory as follows. He generates coherent states according to Alice's description, individually interferes them with the corresponding states in his memory, and checks whether the number of photodetection events at the signal null-port is smaller than $s_a L$, in which case the message is confirmed to be valid. The fraction $s_a$ is called the authentication threshold. Alice must share a well-defined phase reference with all of the receivers to ensure that their measurement of her phase encodings is correct. Our experimental system time multiplexes the encoded signal and delayed reference pulse in one fibre. The exact mechanism will be outlined in the Methods section. Assuming that the authenticity of the message has been confirmed, Bob can then prove to Charlie that he has received that particular signed message from Alice. To do this, he forwards the message and the classical description of the signature states he received from Alice to Charlie. To verify a signed message forwarded by Bob, Charlie follows the same procedure as Bob, but with a modified threshold $s_v > s_a$, called the verification threshold. The difference in the thresholds $s_a$ and $s_v$ is required to ensure security against repudiation by Alice, depends on the parameters of our system and will be given later. Essentially, since Bob and Charlie have symmetric quantum signature states they will obtain the same measurement statistics. The gap between $s_a$ and $s_v$ ensures that Alice cannot make one of them accept and the other reject a message, except with vanishingly small probability.

The choice of mean photon number per pulse $|\alpha|^2$ for the coherent states emitted by Alice to each party depends on the number of possible phase encodings $N$ and the signature length $L$. Figure. 2 shows how the information about the phase of each encoded laser pulse, available to a malicious party, given by the von Neumann entropy[20] of the state $\rho_{\text{Single}}$ defined by:

$$\rho_{\text{Single}} = \frac{1}{N} \sum_{k=0}^{N-1} |\alpha \exp(2\pi i k/N)\rangle \langle \alpha \exp(2\pi i k/N)|, \qquad (1)$$

varies with increasing $|\alpha|^2$, for two receivers ($T = 2$). In practice we must ensure that the information about the whole signature known by Alice far exceeds that which is accessible to a malicious party with access to all signature states in circulation, that is, $L \cdot \log_2(N) \gg L \cdot T \cdot S(\rho_{\text{Single}})$. We will return to the issue of security below; a more detailed analysis is given in the Supplementary Discussion where we find the optimal attacks and calculate the various cheating probabilities. It is an important assumption we adhere to that the channels from Alice to Bob's and Charlie's inputs to the multiport are under the control of the honest receiving parties, that is, an external party may not tamper with the states sent, although the channel is not assumed to be private. Otherwise a man-in-the-middle (impersonation) attack, where the attacker swaps Alice's quantum signatures for their own, becomes possible. If authenticated quantum channels are presumed not to be available impersonation attacks possibly could be countered in a way similar to in quantum key distribution (QKD) since Alice is sending a restricted set of quantum messages. For this reason we have in the more detailed security analysis focused on the aspects of the QDS protocol that are genuinely different from the QKD setting. Schemes for either QKD or QDS that use linearly independent states, and which counter impersonation attacks through partial disclosure of key or signature states and discussion over an authenticated classical channel remain vulnerable to attacks using unambiguous state discrimination[24]. Such attacks, however, only place an upper bound on the total allowed loss. In our implementation, this bound is high and not of concern, since the success probability for an unambiguous measurement that distinguishes between all eight ideal quantum signature states is low, of the order of $10^{-9}$.

**Results**

**Honest Alice:** Figure 4 shows the experimental results obtained by Charlie in the system using eight equally spaced phase encodings ($N=8$) and an honest Alice sending the same signatures to both Bob and Charlie. For these measurements the phase modulator in the Alice to Bob fibre was deactivated. The dashed lines represent the predictions for the quantities using a theoretical model which is explained in more detail in the Methods section while the data points are actual experimentally recorded values. As the mean photon number per pulse launched by Alice into the comparison system increases, so the count-rate at the detectors increases. The multiport null-port count-rates are significantly lower than those at the signal port but are non-zero. The null-port counts are primarily due to the interferometric fringe visibility of the multiport (although background events at the detectors do make a small contribution)[25.] This multiport null-port rate sets a baseline for the system operating with an honest Alice.  The encoding error is defined as the number of temporally filtered pulses detected by a receiver at his signal null-port, divided by the temporally filtered total number of pulses recorded by that receiver.  The encoding error rate is constant within experimental fluctuations across the range of experimentally examined values of $|\alpha|^2$, as the effects of intersymbol interference and background events in the detectors are negligible.  Each of the six detectors has a mean dark-count rate of 320 counts per second, and the probability of temporal intersymbol interference for each detector is $3 \times 10^{-8}$.

**Detection of discrepancies:** A necessary requirement for a system to be immune to Bob's forgery is that Charlie is capable of detecting a discrepancy between Alice's chosen phase encoding in a signature state and Bob's average best guess of the phase.  A discrepancy will cause a higher probability for a photodetection event on the signal null-port in the case when Charlie measures the pulse using a phase different from that actually encoded on the pulse.  We experimentally verified this by looking at the encoding error at Charlie if he measures using a phase different from that defined by Alice.  The results are shown in Fig. 5.  This allows us to characterise the effects of a mismatch between the encodings in true and forged quantum signatures.  The off-diagonal elements correspond to Charlie measuring using a phase different from that used by Alice.  The results show that Charlie can detect an increase in his encoding error percentage when Bob (or another external party) attempts to forge a message.  A greater difference between the probabilities of null-port events for differing and identical phases reduces the required key length for a desired level of security.

Certain types of malicious activities by Alice can also be detected by monitoring the multiport null-port count-rates.  We experimentally tested the case where Alice sends different signatures to Bob and Charlie. The phase modulator in the fibre connecting Alice to Bob was used to change the phase encoding of two pulses in every sixteen by a fixed phase, and Charlie's count-rate at the multiport null-port and error rate were monitored. The results for the raw count-rate at the multiport null-port can be seen in Fig. 6. It can be observed from Fig. 6 that as Alice increases the magnitude of the phase difference between the states, the count-rate at Charlie's null-port increases as expected.

**Discussion**

A more careful security analysis can be found in the Supplementary Discussion, but will be outlined here. We identify two classes of forging attacks. In the *active attack* the malicious party (Bob or Charlie) is allowed to alter the states he forwards to the other party within the multiport to optimise his later cheating probability. In a *passive attack*, the recipients of the quantum signatures are benevolent throughout the signature distribution phase but will attempt to falsify a message later. This is a restricted setting, which corresponds to the case where each recipient is *a priori* equally likely to be the forger, or when a trusted third party holds the multiport. An external party, who does not hold any signature copies, will have a lower probability for successfully forging a message.

The probability of cheating in a passive attack can be evaluated using the experimental results presented in Fig. 5. This probability is also central to estimating the cheating probabilities using active attacks. To counter against active attacks, the outcomes at the multiport null-ports during signature distribution must be taken into account. In short, a low count-rate at the multiport null-port guarantees that the probabilities of cheating using the active and passive attacks will not differ substantially.

We now proceed to calculate the probability of cheating for a passive attack, saving the case of an active attack for the Supplementary Discussion. Assuming Bob is the forger, his optimal passive strategy for forging the message, say $m=0$, consists of producing a "best guess" of the private key by inspecting his copy of the corresponding signature state and forwarding this guess to Charlie. We assume that the phases of the states have been generated independently and uniformly at random. Then Bob's optimal strategy is to employ a single generalised measurement applied on each of the states in his signature. The probability of causing a photodetection event, when verifying a single state in the signature, is then given by

$$p_{\text{forgery}} = \min_{\{\{\Pi_\phi\}\}} \frac{1}{N} \sum_\phi \sum_\theta Tr\left(\Pi_\phi \rho^\theta\right) c_{\phi,\theta}, \qquad (2)$$

where $c_{\phi,\theta}$ is the probability of a photodetection event in Charlie's signal null-port arm, given that $\rho^\theta$ is the coherent state sent by Alice, with phase $\theta$, and the phase angle declared by Bob is $\phi$. The operators $\Pi_\varphi$ describe the measurement made by Bob, on the signature copy or copies he has access to, to select the best possible phase angle $\phi$. Bob's optimal measurement, minimising the probability to cause a photodetection event is a minimum cost measurement[18,] with the cost matrix $C$ with elements $c_{\phi,\theta}$.

The cost matrix $C$ is obtained experimentally for our system, and is related to the encoding error matrix shown in Fig. 5. The cost matrix additionally takes into account vacuum events on the signal ports, which are not included in the calculation of the encoding error. The full cost matrix is given in the Supplementary Discussion. We assume that the states in both Bob's and Charlie's quantum signatures are perfect, without the loss of generality, as (random) imperfections could only degrade the probability of getting Charlie to accept a forgery. If Bob is honest then the maximum probability of causing a photodetection event is $p_{\text{original}}$ (equal to the largest diagonal element(s) of $C$). In our experiment the diagonal elements of $C$ exhibit a small standard deviation around a well defined mean. As long as $p_{\text{forgery}} > p_{\text{original}}$, given a large enough sample size (i.e. signature length $L$), cheating and honest scenarios can be distinguished using statistical methods. For the values $p_{\text{forgery}}$ and $p_{\text{original}}$, one may set the authentication and verification thresholds as $s_a = \frac{1}{3}g + p_{\text{original}}$ and $s_v = \frac{2}{3}g + p_{\text{original}}$. The gap $g = p_{\text{forgery}} - p_{\text{original}}$ appears as the central parameter of the cheating probabilities and is equal to $8.03 \times 10^{-4} \pm 0.3 \times 10^{-4}$ for our system.

The probability of forging using a passive attack equals the probability of a cheating Bob causing fewer than $s_v L$ photodetection events in Charlie's signal null-port arm. Using Hoeffding's inequalities[26,] we bound this as

$$\varepsilon_{\text{forging}} \leq 2\exp\left(-\frac{2}{9}g^2 L\right). \tag{3}$$

Analogously, the probability $\varepsilon_{\text{robustness}}$, for Bob and Charlie to reject a message from Alice, if all parties are honest, is

$$\varepsilon_{\text{robustness}} \leq 2\exp\left(-\frac{2}{9}g^2 L\right). \tag{4}$$

Without errors caused by imperfections in the components, honest Bob and Charlie would never reject a message from an honest Alice. For the derivation of the parameters above see the Supplementary Discussion.

Further to the probability of forging a message, there also exists a probability of repudiation. In order to repudiate her signature, a malevolent Alice needs to prepare the signature states so that Bob accepts and yet Charlie rejects the message when Bob forwards it, or vice versa. For this purpose, she may send different signature states to Charlie and Bob, or more generally, she may use any type of $2L$ mode states, including entangled and mixed states. However, regardless of her choice of states, assuming an ideal multiport, the signature states Charlie and Bob end up with are symmetric under the swap of Charlie and Bob's systems. Thus, the probability matrix describing the *a priori* occurrence of photodetection events in Bob and Charlie is symmetric. To maximise the probability of causing a mismatch in Charlie and Bob, required for repudiation, the most Alice can achieve is that with probability $1/2$, Charlie detects a photon and Bob does not, and with probability $1/2$ the opposite. To ensure repudiation of the signature, the number of cases where Charlie detects photons and Bob does, or vice versa, needs to be higher than $gL$, and the probability of this happening is upper bounded by[10, 16] $\varepsilon_{\text{repudiation}} = (1/2)^{\frac{1}{3}gL}$. Imperfections could increase the factor of $1/2$ in $\varepsilon_{\text{repudiation}}$. However for our system this increase is such that $\varepsilon_{\text{repudiation}}$ is smaller than $\varepsilon_{\text{forging}}$, and thus $\varepsilon_{\text{forging}}$ bounds the overall security of our system.

The work detailed in this article is an experimental demonstration of the distribution of quantum digital signatures. This system could also be used to share quantum frames of reference, and may be applicable to further quantum information protocols and experiments[27, 28]. The current system does not utilise any form of quantum memory and it would be desirable to combine it with some form of the same[23, 36, 37], or find a way to circumvent this requirement completely. Other challenges include extending the distance between Bob and Charlie beyond the current ~5 metres. At present, the pulses from the other party's half of the signature must arrive at the final beamsplitter cube at the same time as those from the party's retained half of the signature. As the distance between Bob and Charlie is increased, so the delay in the paths for the retained portion must increase, and this leads to instabilities increasing the error rate. This could be alleviated by temporarily storing the retained portions in a short-term quantum memory. Currently the system is designed to operate with two receivers and scaling it up to a greater number of receivers is an area for future research. It is possible to generalise the scheme using balanced multiports, as suggested by Andersson *et al.*[16]. Furthermore, Andersson *et al.*[16] suggested that the multiports may be realised in a time-resolved fashion or using fibre couplers.

The primary reason for the small $g$ value is the low photon flux at the receiver's detectors. Reducing the loss of the multiport from the current 7.5 dB would be desirable. It may be possible to replace the air-gaps with fibre stretchers but these can increase the quantity of fibre within the system and lead to reduced up-time. Additionally, reducing the loss of the system overall will reduce the signature length $L$ required for a given level of security. With the current system parameters, for signature lengths $L$ of the order of $5 \times 10^6$ pulses the bounds on the failure probability become non-trivial (and decay exponentially quickly from that point on). Increasing the clock rate, and therefore the transmission rate of the system, is consequently an obvious goal. The phase modulators, lasers and driving electronics, are all capable of clock-rates up to a maximum of 3.3 GHz.

**Methods**

**System implementation:** The experimental implementation of the QDS system is shown schematically in Fig. 3. This system encodes phase onto highly attenuated laser pulses and compares two copies of the quantum signature, simultaneously symmetrising the states, using the multiport. Finally, Bob and Charlie measure the phases of the pulses and detect the photons using silicon single-photon avalanche diodes (Si-SPADs)[29]. The addition of a phase modulator in the final section of fibre between Alice and Bob allows us to test the multiport performance when Alice tries to cheat by sending different signatures to different recipients. The air-gap in the other arm allows the transmission losses to be balanced between each arm so that the same $|\alpha|^2$ value is launched to each recipient, and permits compensation for small path-length differences between the two launch arms; $|\alpha|^2$ is defined after Alice's phase modulator/air-gap at the inputs to the multiport.

To ensure a high interferometric fringe visibility in the interferometers of the system, it is necessary to ensure that the relative path-length differences remain constant to within a fraction of the emission wavelength of the source laser[30]. Adjustable air-gaps in active feedback loops are used to compensate for any slow time dependent variations in the relative path lengths[25]. The fringe visibility is monitored during operation of the system and when a deviation from the expected value is obtained, signature distribution is halted and tuning carried out using a higher intensity signal with known phase modulation until the optimum visibility is obtained. Our interferometers had fringe visibilities of 98%. It is likely that the greatest contributions to the reduction in the visibility of the interferometers from 100% are due to the linewidth of our laser source[30] and loss of polarisation extinction ratio (PER) at non-ideal fusion splice and flat-polish bulkhead joins between the various fibres which comprise the system. The stress members in the polarisation-maintaining fibre must be aligned by eye during the splicing process, introducing human error, and the flat-polish bulkhead connectors have variable misalignment due to manufacturing tolerances.

The system has been assembled from polarisation-maintaining fibre which supports a single mode at a wavelength of 850 nm. The use of polarisation-maintaining fibre ensures good fringe visibility in the interferometers of the system as high visibilities can be achieved when interfering two highly linearly polarised light fields[31,] and the use of a single spatial mode in the fibre reduces temporal broadening of the pulse. An operating wavelength of 850 nm was chosen to provide compatibility with comparatively mature high detection efficiency thick junction Si-SPADs[29]. The system operates at a pulse repetition frequency of 100 MHz to avoid intersymbol interference when using these detectors[29]. Si-SPADs were selected as detectors since the losses of the system (7.5 dB from the comparison stage input to each demodulation interferometer and 7.1 dB for each demodulation interferometer) mean that the pulses transmitted by Alice are in the single-photon regime at the detectors. The detection efficiency of the Si-SAPDs used for these experiments exhibits a count rate dependent variation. At higher count rates, the detection efficiency of the detectors decreases, reaching a minimum of 36.8% as opposed to the maximum value[32] of 40%.

This system time multiplexes a phase reference pulse between successive 100 MHz clocked signal pulses using an asymmetric double Mach-Zehnder approach as used in many quantum key distribution systems employing phase basis sets[33.] In an ideal system, the receivers would utilise their paths with air-gaps to delay only the signal pulse so that it recombines with the corresponding reference, revealing the phase encoding. However, in a real system there will be photons which take non-interfering paths in sender and receiver (i.e. both short paths or both delayed paths) contributing nothing to the signature[33] and these are software gated from the photon arrival times recorded using the free-running Si-SPADs. In post-processing, the time gating software opens a window of duration 2 ns centred on the expected arrival time of a pulse and disregards events which occur outside of this window.

**Photon source characterisation:** A vertical cavity surface emitting laser (VCSEL) emitting at a wavelength of 849.8 nm, and with a spectral full-width at half maximum (FWHM) of 0.23 nm, was selected as the photon source in these experiments. In common with most other diode lasers, VCSELs exhibit a temperature dependent output wavelength[35] and consequently the VCSEL used in these experiments was mounted on a custom temperature controller and maintained at an operating temperature of 15±0.1 °C to ensure wavelength stability. The central wavelength of the laser had a measured wavelength shift of 77 pm/°C. The laser was driven using a 500 ps wide square electrical pulse at a repetition rate of 100 MHz using a commercial driving board which ultimately produced optical output pulses of duration 780 ps FWHM.

To improve the PER of the VCSEL[35,] it was necessary for the laser output to be transmitted through two high extinction ratio (in excess of 10,000:1) polarisers before being launched into the single-mode polarisation maintaining fibre from which the main part of the optical system was comprised. The VCSEL output had a PER of 7:2 while for the light after the cleanup polarisers the PER was measured as being in excess of 1,200:1.

Our measurement of the pulse-to-pulse variance of the output power of the VCSEL was limited by the resolution and noise level of our detector, but can be stated to be lower than 3%. The mean photon number per pulse $|\alpha|^2$ was set using a computer-controlled motorised attenuator. A stepper motor drives a screw into or out of a collimated beam to provide the required attenuation. The motorised attenuator exhibits reproducibility of attenuation setting to within 1% of the calibrated value. In all uncertainty analyses we have assumed a worst case scenario that the uncertainty in $|\alpha|^2$ is dominated by the pulse-to-pulse variance in the output power of the VCSEL.

The sender and receivers utilise phase modulators with a voltage to enact an optical phase change of $\pi$ radians ($V_\pi$) of 6 V. The driving electronics for the phase modulator have a pulse to pulse amplitude variance which corresponds to a variance in the desired phase encoding of $\pm 1.6 \times 10^{-3}$ radians or $\pm 0.2\%$ of the difference between successive values when 8 encodings are used.

**Theoretical modelling:** The theoretical model for the count-rates is based on our previous work detailed in ref. *25* The theoretical model requires knowledge of the system losses, clock rate (100 MHz), detector dark count rate (320 counts per second), detector detection efficiency (42%), classical visibility (98%) and system instrument response function (modelled as the PerkinElmer thick junction silicon single photon avalanche diodes of per ref *25*). To calculate the raw count-rate at the receiver we model the multiport and receiver as losses of 7.5 dB and 7.1 dB respectively, without taking into account interferometric visibility, as in equation 1 of ref. *25*. The temporally filtered count-rate at the receiver was calculated following the same method as outlined in ref. *25* with the same instrument response function parameters. The encoding error was calculated using a modified form of the equations used to predict the quantum bit error rate (QBER). In equation 8 of ref. *25* the protocol-dependent scaling term ($\alpha_{\text{Protocol}}$) was set equal to unity so that the dark count contribution was given by

$$\frac{(1/2)\upsilon R_{\text{Dark}} \Delta T}{R_{\text{TimeGated}}(\Delta T)} \tag{5}$$

where $\upsilon$ is the clock frequency of the system (100 MHz), $R_{\text{Dark}}$ is the detector dark count rate (320 counts per second), $\Delta T$ is the time gate duration (2 ns centred on the expected peak position) and $R_{\text{TimeGated}}(\Delta T)$ is the count rate remaining after temporal filtering by the 2 ns gate, as indicated by the triangular points on Figure 4. Calculation of the temporally filtered rates proceeded as in the case of ref. *25*.

The raw count rates at the null, or vacuum state, ports of the multiport, were predicted by modelling the multiport as a loss of 7.5 dB and then utilising the definition of visibility as

$$Visibility = \frac{I_{\text{Max}} - I_{\text{Min}}}{I_{\text{Max}} + I_{\text{Min}}} \quad (6)$$

(where $I_{\text{Max}}$ is the intensity of an interference maximum and $I_{\text{Min}}$ is the intensity of an interference minimum) with the count rates substituted for the intensities. For a visibility of 98% the theoretically predicted count rate at the signal output was substituted into the visibility equation as $I_{\text{Max}}$ and the equation rearranged to give $I_{\text{Min}}$ as the count rate on the multiport null port.

The theoretical model can be used to predict the results shown in Fig. 5 (and by extension the cost matrix $C$) and Fig. 6. The visibility is modelled using equation 6 above, equation 5 of ref *25* and the cosine dependency of the phase sensitivity of a Mach-Zehnder interferometer given in equation 36 of ref 38.

**Acknowledgments:** This work was supported by the UK Engineering and Physical Sciences Research Council (EPSRC) under EP/G009821/1. V.D. is also affiliated with Ruđer Bošković Institute, Zagreb, Croatia.

**Author Contributions:** R.J.C. and P.J.C. designed and assembled the experimental system, collected and analysed the experimental results and developed the theoretical model for the count-rates and encoding errors. V.D. carried out the security analysis with assistance from J.J.

and suggested experiments to be performed. E.A. supervised the security analysis and co-authored the original theoretical paper on the quantum digital signature scheme using coherent states which has been experimentally implemented in this work. G.S.B, E.A. and J.J. secured funding. G.S.B. acted in an overall supervisory role. All authors contributed to the submitted manuscript which is based on an initial draft by R.J.C. and V.D..

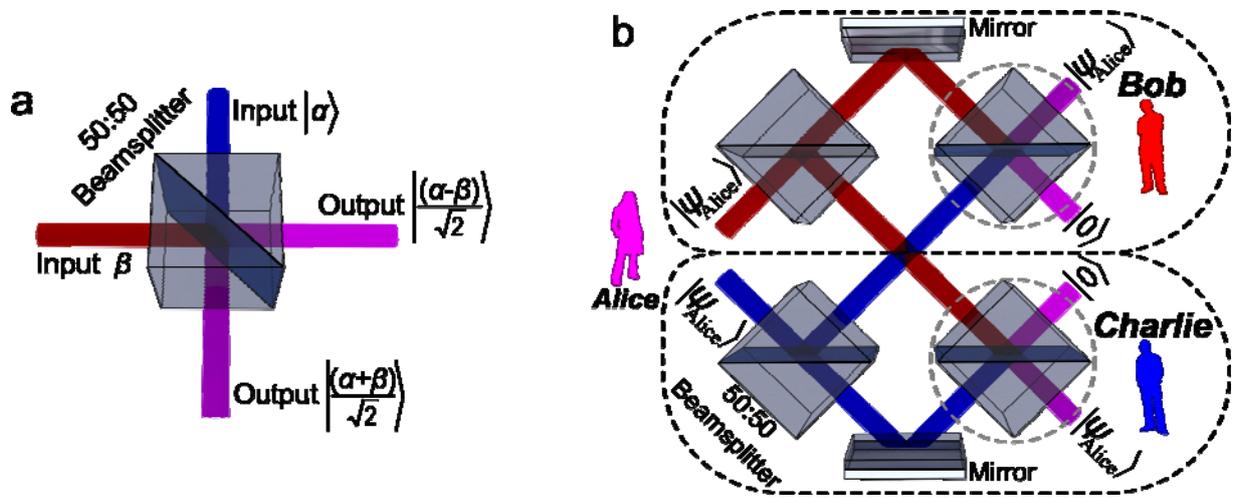

**Figure 1 | The basic principles of our quantum digital signature scheme.** (a) Field mixing on a 50:50 beamsplitter cube. The states $|\alpha\rangle$ and $|\beta\rangle$ are coherent states[15, 17]. (b) Quantum digital signature distribution in the case where Alice (the sender) is not trusted. Bob (one of the receivers) separates half of each signature state received from Alice using a beam splitter, and sends it to Charlie, where it is compared with the signature half Charlie received from Alice using the 2$^{nd}$ beam splitter within the grey dashed lines. Likewise, Bob compares the signature state he received directly from Alice with the one received via Charlie. The final beamsplitters in the dashed grey circles work as depicted in Fig. 1(a).

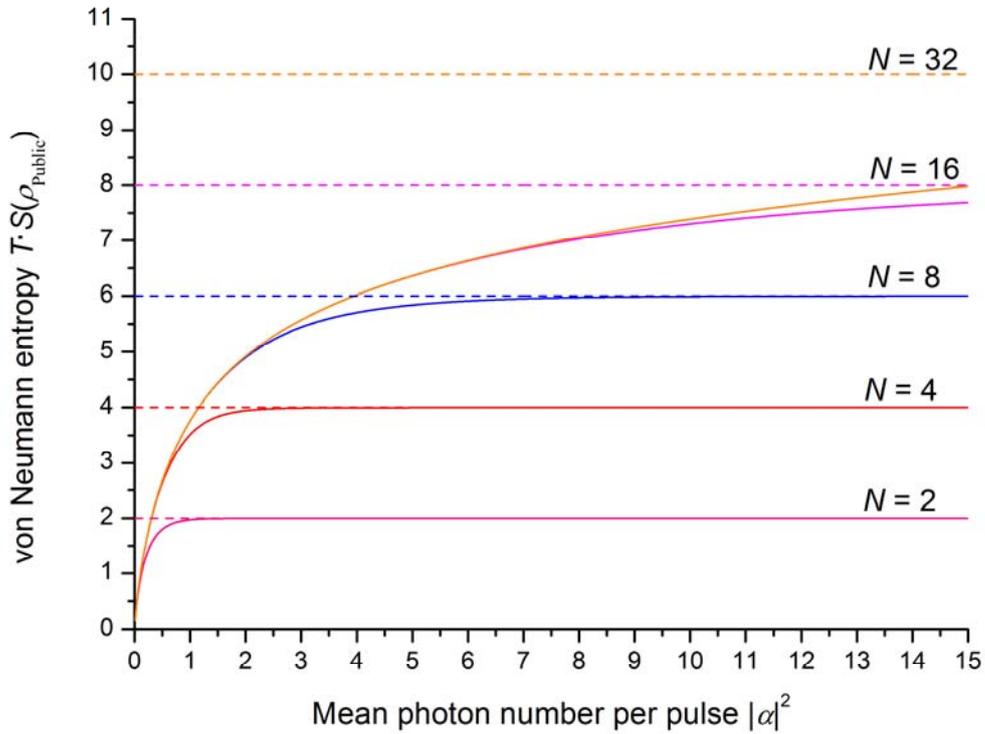

**Figure 2 │ The dependence of the von Neumann entropy for two receivers on mean photon number.** The von Neumann entropy $S(\rho_{Single})$ is shown scaled for two receivers ($T = 2$). Five different possible numbers of phase encodings $N$ equal to 2, 4, 8, 16 and 32 are shown. The asymptotic value of the entropy increases with the number of phase encodings $N$, indicating that it is possible to use higher mean photon numbers $|\alpha|^2$ for greater values of $N$.

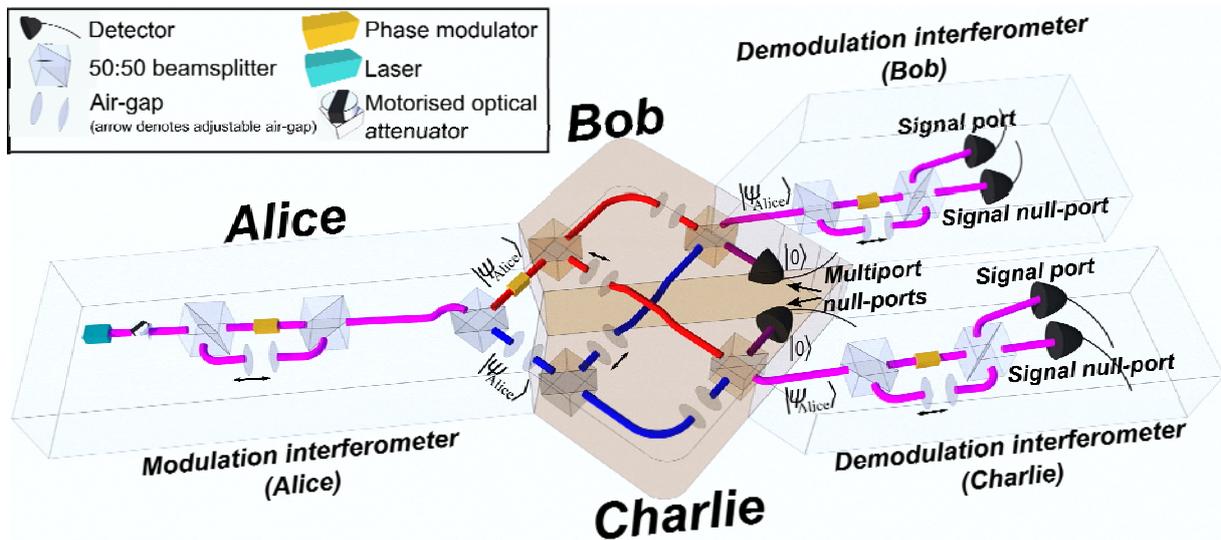

**Figure 3│ A schematic diagram of the fibre-based experimental demonstration of quantum digital signatures**. The laser used is a vertical cavity surface emitting laser (VCSEL) and the detectors are single-photon avalanche diodes (SPADs). The system is constructed from polarisation maintaining fibre to improve interferometric fringe visibility. The final phase modulator within Alice's apparatus and corresponding adjustable air-gap can be used to test certain malicious activities implemented by Alice and are removed for experiments with honest parties.

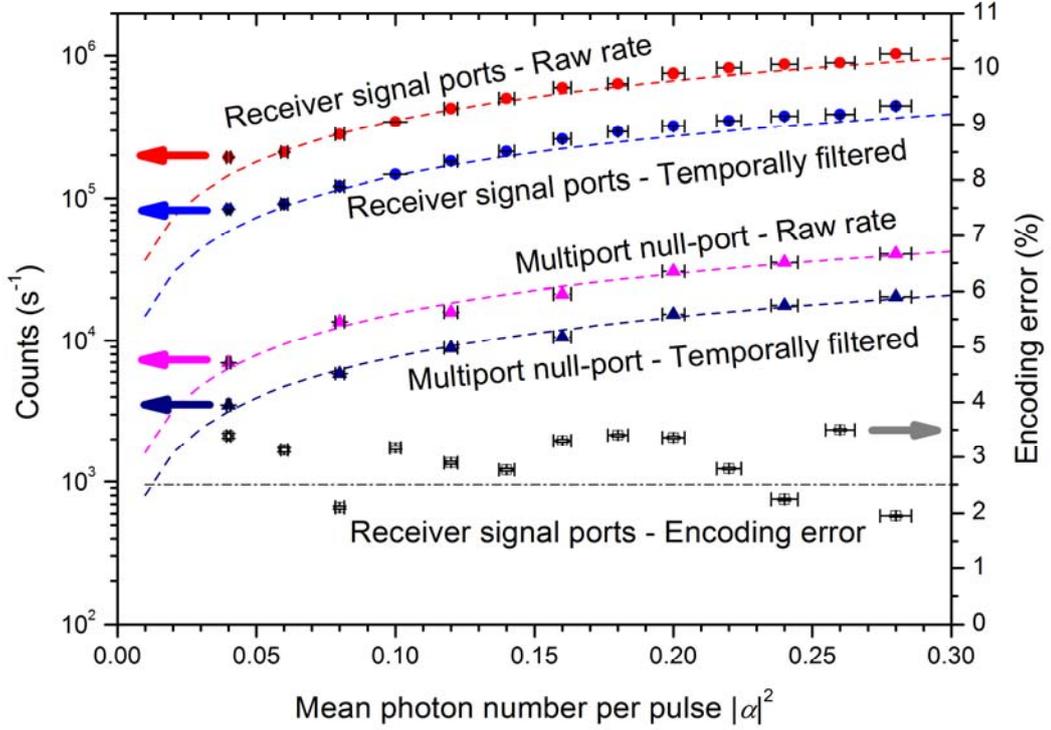

**Figure 4 | Experimentally measured detector event count rates and encoding error for receiver Charlie.** The system clock rate was 100 MHz and eight equally spaced phase encodings were used. Data points represent actual experimental results while dashed lines are theoretical predictions. The raw count-rate is the detector click rate summed over both of Charlie's signal SPADs after the demodulating interferometers. The time gated count-rate is the raw count-rate after temporal filtering using a 2 ns duration window centred on the expected arrival time to reduce the effects of background events, temporal intersymbol interference[25] and non-interfering photons. The encoding error is the number of temporally filtered detector events recorded by Charlie at the signal null-port divided by the total number of temporally filtered detector events he recorded. The experimental values have a square root uncertainty in count rate, while uncertainty in the mean photon number is dominated by a worst case scenario

assumption than the pulse-to-pulse variance in the output power of our laser is the experimentally measured maximum of ±1.5%.

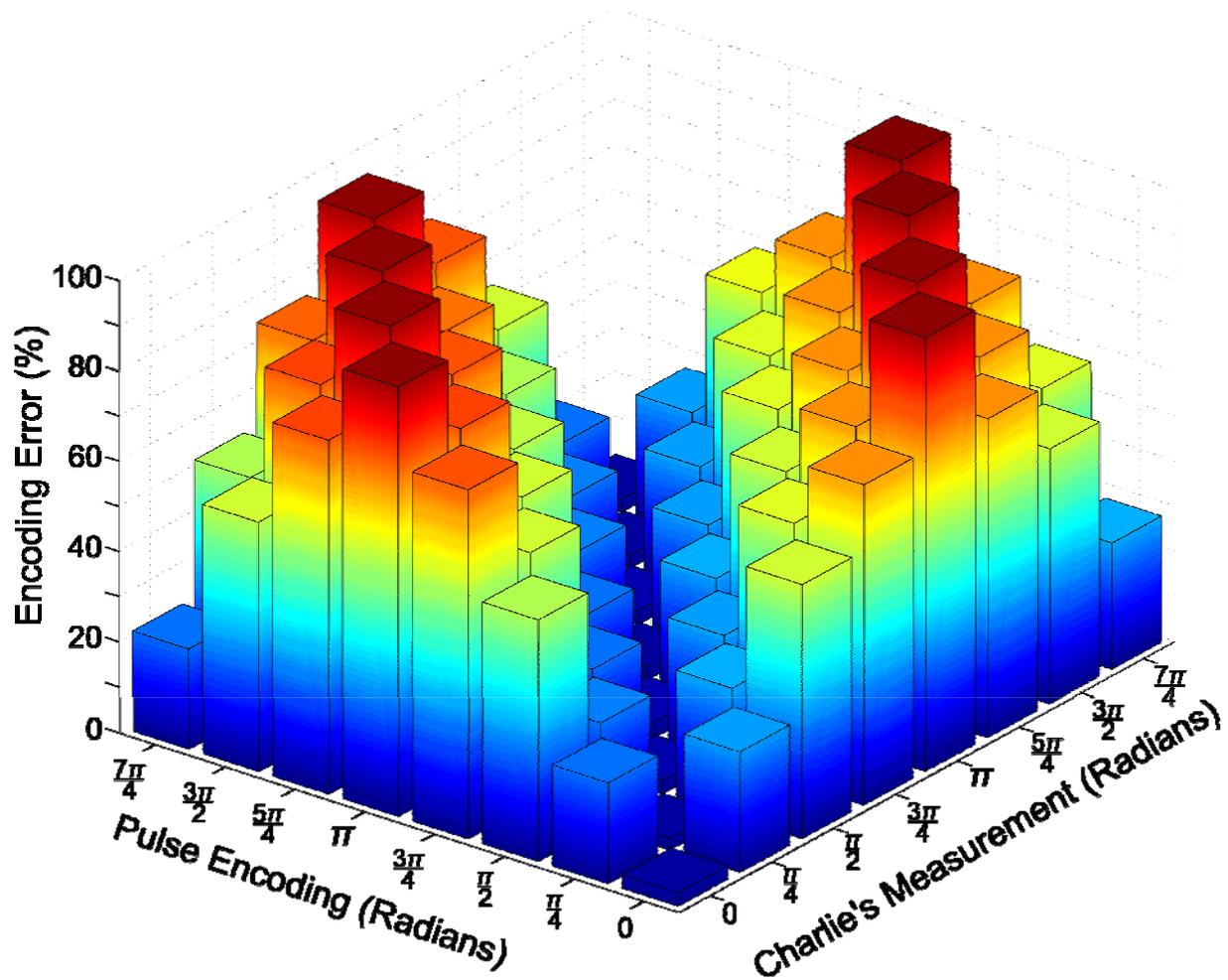

**Figure 5│ The percentage encoding error for Charlie when he measures using states with phases identical and different from those defined by Alice.** The coherent states have mean photon number $|\alpha|^2 = 0.16$ and are chosen from a set of $N = 8$.

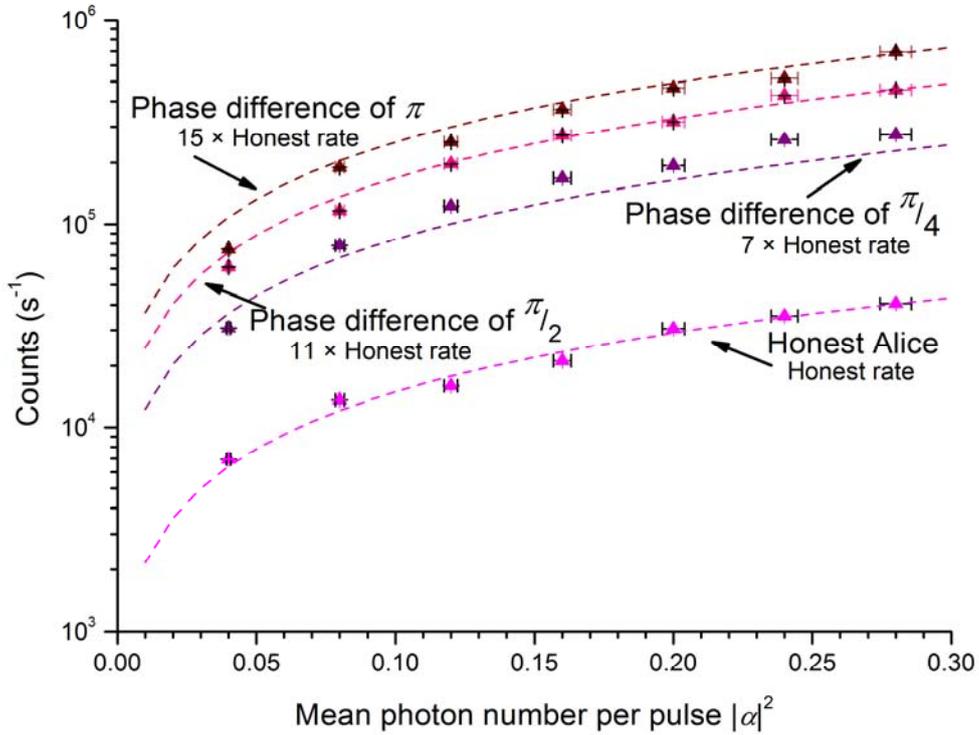

**Figure 6 | The effect on the multiport null-port count rate if Alice sends different signature states to Bob and Charlie.** Data points represent actual experimental results while dashed lines are theoretical predictions. She alters the phase encoding of two pulses in every sixteen by a fixed amount. The "Honest rate" is the observed multiport null-port count-rate when Alice sends the same signature to Bob and Charlie. As Alice makes progressively greater changes to the phase encodings sent to one party, the count-rate at the multiport null-port increases. It can be seen that a phase difference of $\pi/4$ increases the multiport null-port count rate by a mean factor of 7, a phase difference of $\pi/2$ increases it by a mean factor of 11 and a phase difference of $\pi$ increases it by a mean factor of 15. The experimental values have a square root uncertainty in count rate, while uncertainty in the mean photon number is dominated by a worst case scenario

assumption than the pulse-to-pulse variance in the output power of our laser is the experimentally measured maximum of ±1.5%.

# Supplementary Information for
*Experimental demonstration of quantum digital signatures using phase-encoded coherent states of light*

Patrick J. Clarke[1], Robert J. Collins[1], Vedran Dunjko[1], Erika Andersson[1], John Jeffers[2], Gerald S. Buller[1]

[1]SUPA, School of Engineering and Physical Sciences, David Brewster Building, Heriot-Watt University, Edinburgh, EH14 4AS, UK.
[2]SUPA, Department of Physics, John Anderson Building, University of Strathclyde, 107 Rottenrow, Glasgow, G4 0NG, UK.

Correspondence to: r.j.collins@hw.ac.uk.

# Supplementary Figures

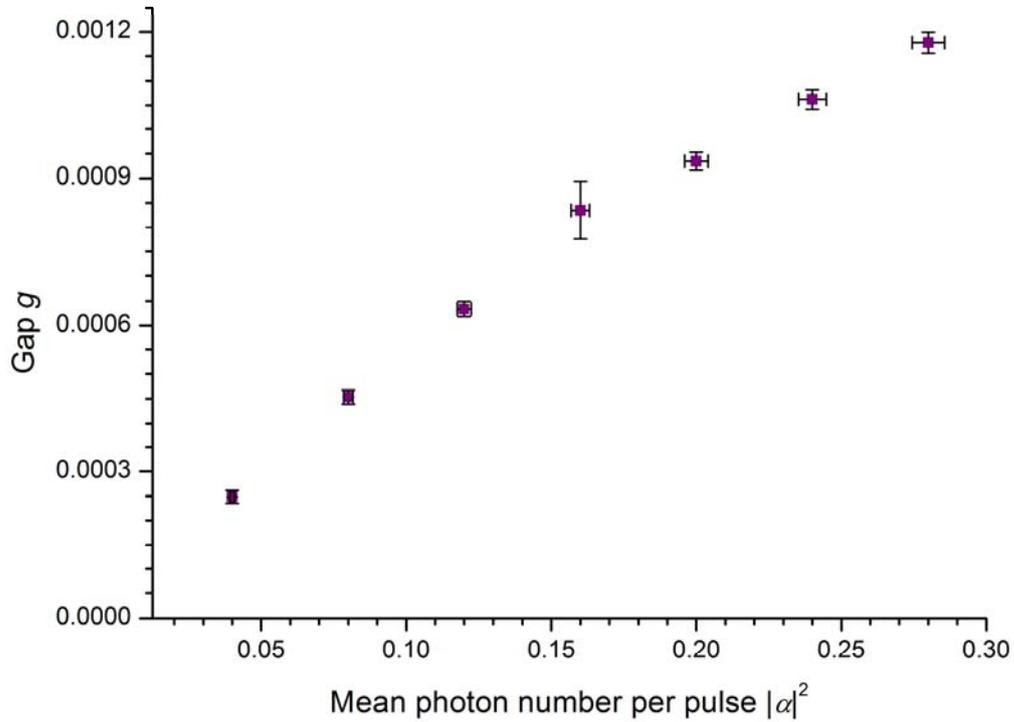

**Supplementary Figure S1│An estimation of the variation of the gap with mean photon number per pulse.** This estimation is based on predictions for the cost matrices and count rates that have been produced using the theory model described in the Methods section To calculate the errors we considered that there is a square root uncertainty in count rate, while uncertainty in the mean photon number is dominated by a worst case scenario assumption than the pulse-to-pulse variance in the output power of our laser is the experimentally measured maximum of ±1.5%.

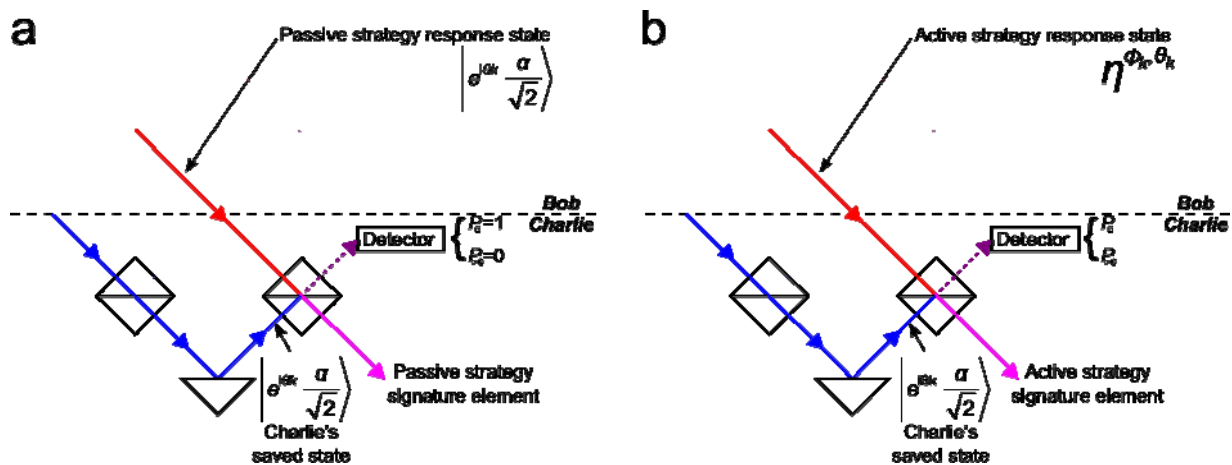

**Supplementary Figure S2 | Passive and active cheating strategies in the multiport.** (a) In the passive strategy, Bob's response state is equal to Charlie's saved state. Consequently, the probability $P_0$ of not detecting a photon at the multiport null-port is unity whereas the probability of detecting one or more photons $P_{>0}$ is zero. (b) In an active strategy, the probability $P_0$ is non-zero. It is, however, equal to the expected fidelity between the passive and active strategy signature elements.

## Supplementary Table

|  | Charlie multiport signal (counts s$^{-1}$) | Bob multiport signal (counts s$^{-1}$) | Charlie multiport null-port (counts s$^{-1}$) | Bob multiport null-port (counts s$^{-1}$) |
|---|---|---|---|---|
| Multiport input to Charlie blocked | $9.1\times10^5$ | $2.84\times10^5$ | $9.6\times10^5$ | $2.61\times10^5$ |
| Multiport input to Bob blocked | $1.01\times10^6$ | $6.25\times10^5$ | $8.5\times10^5$ | $7.17\times10^5$ |

**Supplementary Table S1 │ The count rate of the multiport outputs were monitored when alternatively one of the inputs to the multiport, either at Bob or at Charlie, were blocked.** All values were measured using the same thick junction silicon avalanche diode.

## Supplementary Discussion

In this section we provide the details of the security analysis of the three-party QDS protocol realised using coherent states. The security analysis assuming passive attacks is based on observed experimental results, whereas the preliminary analysis of security under general types of attacks is performed under certain assumptions. In particular, we calculate the probabilities of successful forging, and repudiation by malevolent parties as functions of private key lengths, and show that the presented protocol is asymptotically robust.

**Protocol outline**

1. To sign a single bit (message $m = 0$ or $1$) in the future, Alice generates two sequences $PrivKey_0 = \left(\theta_1^0, \ldots, \theta_L^0\right)$ and $PrivKey_1 = \left(\theta_1^1, \ldots, \theta_L^1\right)$, of $L$ randomly chosen angles from the set of $N$ equally spaced phases, so $\theta_k^m \in \{\frac{2r\pi}{N} \mid r = 0, \ldots, N-1\}$. The pair $\left(m, PrivKey_m\right)$ is called a private key pair for message $m$.

2. Alice then generates two copies of a sequence of coherent states $QuantSig_0 = \left(\rho_1^0, \ldots, \rho_L^0\right)_{k=1}^{L}$ with the coherent phases matching the angles in the sequence $PrivKey_0$, thus $\rho_k^0 = |e^{i\theta_k^0}\alpha\rangle\langle e^{i\theta_k^0}\alpha|$ where $\alpha$ is a real positive amplitude. A sequence of such states is called a quantum signature. She sends a copy of the quantum signature to each of Bob and Charlie each, informing them that they correspond to message $m = 0$. Alice then does analogously for the message $m = 1$. The individual state $\rho_k^m$ we refer to as the $k^{th}$ quantum signature element state for message $m$.

3. Bob and Charlie send their copies of the sequences $QuantSig_0$ and $QuantSig_1$ through the multiport, saving the output states in quantum memory, noting which quantum signature corresponds to message $m = 0$ and which to $m = 1$. The exit null-ports on Bob's and Charlie's side of the multiport are equipped with photon detectors and the total number of photon events here will serve to disable certain types of forging attacks, but are not crucial for security against message repudiation. For the simple case of passive attacks which we define and analyse first, these outcomes will be ignored.

4. To sign a single bit, say $m = 0$ with Bob, Alice announces the message $m$ and the corresponding private key to Bob (thus she sends the pair $\left(0, PrivKey_0\right)$ over an untrusted channel). To authenticate the signature, Bob generates coherent states of amplitude $\alpha$ with the relative phase defined by the declared private key, and interferes them individually with the states he has in his quantum memory. He monitors the number of photodetection events on his signal null-port arm and confirms the authenticity of the message (*i.e.* the *message passes authentication*) if the number of photodetection events was below $s_a L$. The parameter $s_a$ is called the **authentication threshold**.

5. To prove to Charlie that he received the message $m = 0$ from Alice, Bob forwards to Charlie the pair $(0, PrivKey_0)$ he received from Alice. Charlie then performs an analogous procedure to Bob, and he verifies the message (*i.e.* the *message passes verification*) if his number of photodetection events is below $s_v L$ where $s_v$ is called **the verification threshold**, with $0 < s_a < s_v < 1$.

If any of the thresholds are breached, the protocol is aborted.

**Definitions of security**

The presented Quantum Digital Signatures protocol is designed to be immune to two types of malicious activities: forging and repudiation. Immunity to forging signifies that any receiving party will reject any message which was not sent by Alice herself. Immunity to repudiation signifies that if Alice sends a message to Bob which passes authentication, afterwards the same message will pass verification with Charlie as well – *i.e.* Alice cannot make Bob and Charlie disagree on the authenticity (and consequently the content) of her message.

More formally we have the following:

- We say that <u>a protocol realising QDS is secure against forging</u> if the probability of a recipient successfully producing, without receiving it from Alice, a private key of message *m* which will pass verification by the other recipients is *decaying exponentially quickly in terms of the quantum signature length $L$*.

- We say that <u>a protocol realising QDS is secure against repudiation</u> if, for any malicious activity on the side of Alice, the probability of a message failing verification with one recipient once it has already passed authentication with the other is *decaying exponentially quickly in terms of the quantum signature length $L$*.

- We say that <u>a protocol realising QDS is robust</u> if in the setting where all parties are honest, a message will be authenticated and verified *except with probability decaying exponentially quickly in terms of the quantum signature length $L$*.

Throughout this document we will always consider Bob to be the forger. Note that any security can only be guaranteed if only one party is cheating - two cooperating parties can always cheat on the third. Thus, when analysing security against forging, Alice is assumed to be honest, and in the security against repudiation, Bob and Charlie are assumed to be honest. We also assume that the quantum channel from Alice to an individual recipient is under the recipient's control during the distribution step (step 2 in the protocol outline). This means that while the quantum channel is not assumed to be private, Alice and the recipient have some means to ensure that an external party is not tampering with the states sent over this channel. Authenticity of public keys is the usual assumption in public key cryptographic schemes. The quantum signatures in QDS bare resemblance to public keys in classical cryptography however given that they are quantum states this parallel is not perfect. If one allows public keys to be quantum states then the assumption on the authenticity of the quantum channels to the multiport is simply the standard assumption in public key settings. Otherwise some method of authenticating quantum signatures has to be introduced. A generic solution to this problem would be to employ some type of quantum message authentication scheme[39] which would have to be adapted to work for coherent states. However since the set of messages Alice sends are restricted and similar to states used in QKD it is possible that more direct approaches to resolving a man-in-the-middle-attack (attack by impersonation) like the ones discussed in the main paper may be employed. This could possibly be achieved through means similar to those used in quantum key distribution, noting that the security for this has only been rigorously proven in the qubit setting. That is, by Alice disclosing some of the signature states, the recipient measuring the states accordingly, and checking for discrepancies in Alice's description of the states and the measurement results. As for quantum key distribution, the required exchange of classical information is then assumed to take place over classical authenticated channels. Since this step is intuitively analogous to the situation for quantum key distribution, we will concentrate our security analysis to aspects which are different to quantum key distribution, namely message transferability and forging by a signature recipient. Similarly, it is assumed that Alice cannot tamper with the classical channels used in the verification step (step 5 in the protocol outline). This authentication could be ensured by classical private key authentication schemes, or by employing the QDS protocol itself, which also ensures message authentication.

Related to this, an external party with access to the quantum channels during signature distribution could attempt a measure-and-resend attack. If linearly independent states are used, then both for quantum key distribution and quantum digital signatures, the success probability of unambiguously identifying the state will limit the total loss between sender and recipient that can be tolerated[22]. In our case, however, the tolerated loss is relatively high, since the probability to unambiguously identify the correct state among the eight states used is considerably lower than for standard quantum key distribution schemes using two non-orthogonal states. The requirement stated in the main paper that the accessible information in the $N$ signature should be much lower than $\log N$ will also contribute to the fact that the probability to correctly identify the state must necessarily be low. We also note that we have implemented a first experimental realisation of a quantum digital signature scheme, and we do not rule out realisations of protocols employing linearly dependent states, so that this type of attack is impossible.

**Security against repudiation - Cheating Alice**

The formal definition of security against message repudiation is given in terms of a conditional statement: if one recipient party authenticates the message (say Bob), the other party (Charlie) will verify it as well. This definition agrees with the initial security requirements given in ref *10* and ref *16* and is interpreted as a guarantee to the recipients in the protocol.

Thus, assuming Alice wishes to cheat on Bob, Alice succeeds in cheating only if she gets Bob to accept, and Charlie to reject. This means that she sends a message, say $(0, PrivKey_0)$, to Bob, he checks it and forwards it to Charlie, who then rejects. For the remainder of this section we analyse the probability of this happening. The robustness of the protocol shall be addressed later on in this document.

The most general state Alice can prepare is $\pi_{A,B_1,C_1,B_2,C_2,...,B_L,C_L}$, which is a general $2L+1$-partite state. Subsystem $A$ Alice keeps, and sends partitions $B_1,...B_L$ to Bob and $C_1,...C_L$ to Charlie. If Alice is honest there is no subsystem $A$, and $C_i$ and $B_i$ are identical coherent states with a complex phase known to Alice alone, as specified by the protocol.

**Cheating Alice - separable attack**

We first assume that the multiport Bob and Charlie have is ideal and that the system $A$ is disentangled from the rest of Alice's state (or simply does not exist), and the subsystems $(B_k C_k)$ and $(B_l C_l)$ are not entangled with each other for $l \neq k$. However, we allow the partitions $B_k$ and $C_k$ to be mutually entangled. This type of an attack we refer to as a separable attack. According to the protocol specifications Charlie and Bob will individually run the pairs of states in the systems $(B_k C_k)$ through the multiport, and commit to quantum memory whatever comes out at their signal outputs of the multiport. For the purposes of showing security against repudiation, we can assume that they ignore the measurement outcomes on the multiport null-ports.

For the $k^{\text{th}}$ signature element, the joint system of Charlie and Bob which they store into memory is some state $\pi^{\text{out}}_{B_k C_k}$ which is symmetric under permutations of Bob's and Charlie's subsystems, as we now show. Let

$$\pi^{\text{in}}_{B_k C_k} = \int_{\square^2} P(\alpha, \beta) |\alpha\rangle\langle\alpha| \otimes |\beta\rangle\langle\beta| \, d^2\alpha \, d^2\beta \tag{S1}$$

be any general two mode state given in the $P$ representation. Then the stored output state (when the null-port subsystems have been traced out) is

$$\pi^{\text{out}}_{B_k C_k} = \int_{\square^2} P(\alpha, \beta) |(\alpha+\beta)/\sqrt{2}\rangle\langle(\alpha+\beta)/\sqrt{2}| \otimes |(\alpha+\beta)/\sqrt{2}\rangle\langle(\alpha+\beta)/\sqrt{2}| \, d^2\alpha \, d^2\beta, \tag{S2}$$

which is symmetric in the sense given above. In the process of checking Alice's message, Bob and Charlie will perform a sequence of measurements on their subsystems, and the measurements will be identical as they are prescribed by the (same) private key Alice had sent. Since the systems Bob and Charlie have are symmetric under the swap of their subsystems, the *a priori* probability matrix of their respective outcomes will be symmetric as well.

To explain this let us focus on the $k^{\text{th}}$ subsystem $\pi^{\text{out}}_{B_k C_k}$ given above. Let Bob be the first to check this subsystem, by preparing some coherent state prescribed by the (common) private key Alice declared, and interfering that state with his corresponding element of the quantum signature, namely $Tr_{C_k}\left(\pi^{\text{out}}_{B_k C_k}\right)$, checking whether he gets a photodetection event on his signal null-port. This constitutes a two-outcome measurement characterised by POVM elements $\Pi_0$ and $\Pi_1$ corresponding to registering a photodetection event or not, applied on the state $Tr_{C_k}\left(\pi^{\text{out}}_{B_k C_k}\right)$. If Charlie was the first to check, he would have done the same measurement on the subsystem $Tr_{C_k}\left(\pi^{\text{out}}_{B_k C_k}\right)$. But, since $\pi^{\text{out}}_{B_k C_k}$ is symmetric under subsystem swap, the probability matrix of the joint four outcome measurement $\Pi_{i,j} = \Pi_i \otimes \Pi_j$, for $i, j = 0, 1$, is symmetric as well, so for every possible state $\pi^{\text{out}}_{B_k C_k}$ the probability of getting event outcomes $(0,1)$ (only Charlie registers a photodetection event) and $(1,0)$ (only Bob registers a photodetection event) is the same. Assume Alice wishes Bob to accept and Charlie to reject. Alice requires Charlie to accumulate more photodetection events than Bob. Then the *a priori* probability of Bob not detecting any photons, and Charlie detecting one or more photons, is no higher than $1/2$ (as the opposite event must be equally likely). According to the protocol specifications, Bob accepts if he gets less than $s_a L$ photodetection events. Charlie accepts at less than $s_v L$ photodetection events. Thus, Charlie needs to accumulate $(s_v - s_a) L$ photodetection events more than Bob in order for Alice's cheating to succeed. The choice of values of $s_a$ and $s_v$ come from the security analysis against forging and will be calculated later. The probability of Alice achieving her goal of getting, say Bob to accept, and Charlie to reject is then $(d)^{(s_v - s_a)L}$, as was shown in ref *16*, where $d$ is the probability of getting the outcome $(0,1)$ and equally $(1,0)$. This is maximised for the highest allowable $d = 1/2$ yielding an overall probability of Alice cheating successfully as

$$\varepsilon_{\text{repudiate}} = \left(\frac{1}{2}\right)^{(s_v - s_a)L}. \tag{S3}$$

**Security against repudiation – coherent attacks**

In a coherent attack, the entanglement of the states Alice may use is unrestricted. The probability of Alice cheating, in the case where the entanglement of the states she uses to cheat is restricted as described in the previous section, does not change if we slightly modify the protocol: Alice first sends the elements of the quantum signatures, Bob and Charlie run them through the multiport, Alice sends the corresponding angle to both Charlie and Bob, they immediately measure, and at each step Alice learns the outcomes of Bob's and Charlie's measurements. This modification does not increase a malevolent Alice's cheating probability. Thus we have

$$P(\text{Alice Cheats} \mid \text{original protocol, separable attack}) = \tag{S4}$$
$$P(\text{Alice Cheats} \mid \text{modified protocol, separable attack}),$$

and we will next prove

$$P(\text{Alice Cheats} \mid \text{modified protocol, separable attack}) = \tag{S5}$$
$$P(\text{Alice Cheats} \mid \text{modified protocol, coherent attack}).$$

Finally, we will show that the modified protocol can only help Alice in the coherent attack, leading to

$$P(\text{Alice Cheats} \mid \text{modified protocol, coherent attack}) \geq \tag{S6}$$
$$P(\text{Alice Cheats} \mid \text{original protocol, coherent attack}).$$

This proves that coherent attacks cannot help Alice.

As noted, the most general state Alice could use in her attempt to cheat is $\pi_{A,B_1,C_1,B_2,C_2,\ldots,B_L,C_L}$. The subsystem $A$ remains with Alice, and the rest is sent to Bob and Charlie and will traverse the multiport. First of all, note that in the original protocol, there is no interaction between Alice on one side and Bob and Charlie on the other, once she has declared her private key. If Alice had a system $A$ which is still entangled with whatever Bob and Charlie save after the multiport action, the action of measurement by Bob and Charlie in the verification part cannot convey any information to Alice through the system $A$, since she does not learn the outcomes of Bob's and Charlie's measurements. Hence, she cannot gain anything by manipulating system $A$, and in the original protocol we may assume that Alice simply uses the state

S9

$$Tr_A(\pi_{A,B_1,C_1,B_2,C_2,\ldots,B_L,C_L}). \tag{S7}$$

We will now show that if she uses a separable strategy in the modified protocol, Alice can achieve the same measurement statistics during verification and authentication as by using a coherent attack. Initially, we assume an ideal multiport. The first state Bob and Charlie may measure is $Tr_{B_2,C_2,\ldots,B_M,C_M}(M(\pi_{B_1,C_1,B_2,C_2,\ldots,B_L,C_L}))$ where $M$ denotes the global action of the multiport. Alice could simply have sent this state to Charlie and Bob and achieved the same measurement statistics as for this state (any state she sends which is already symmetric will not be changed by the multiport). However, the measurement outcome may influence the rest of the system, which Alice has not yet sent to Bob and Charlie. However, if the authentication and verification measurement outcomes are revealed to Alice at each step then the state of the rest of her system is also known to her at each step. Then in the sequential setting, she can prepare the corresponding signature state for the second measurement and attain the same measurement statistics. This continues inductively.

Thus we have shown the following: any measurement statistics achieved using a globally entangled cheating state can be achieved using a separable attack, if Alice is allowed to learn the measurement outcomes before sending the next pair of states.
This proves the required claim

$$P(\text{Alice Cheats} \mid \text{modified protocol, individual attack}) = \tag{S8}$$
$$P(\text{Alice Cheats} \mid \text{modified protocol, coherent attack}).$$

To finalise our proof we need to show that
$$P(\text{Alice Cheats} \mid \text{modified protocol, coherent attack}) \geq \tag{S9}$$
$$P(\text{Alice Cheats} \mid \text{original protocol, coherent attack}).$$

This is easy to see as by simply ignoring the information Alice additionally gets in the modified protocol, what Alice runs is effectively the original protocol, barring the timing of the measurements. However, the timing cannot influence the measurement statistics, and hence cannot influence Alice's cheating probability. So our claim holds and using globally entangled states cannot help Alice repudiate her signed messages.

To summarise, as long as the properties of robustness and security against forging can be maintained for some $s_v$ strictly greater than $s_a$, then security against repudiation can be guaranteed as well.

**Security against repudiation with realistic devices**

In the security analysis against repudiation for the ideal case the crux of the argument is that the states Bob and Charlie share are symmetric under swap of their respective subsystems, as this guarantees that for a single pair of states, the probabilities of the outcomes $(0,1)$ and $(1,0)$ of joint measurements made by Bob and Charlie are equal. Since these are equal, each is at most 1/2, and this value is raised to the exponent $(s_v - s_a)L$ to obtain the upper bound on the probability of Alice successfully cheating. Here we briefly address the effects imperfect realisation may have on the security of our system. Note that the multiport acts as a CPTP map (completely positive trace preserving map, quantum channel) on the input state, where the output state is the joint state of the Bob's and Charlie's multiport signal outputs, i.e. elements of the quantum signature. Let $M_{\text{ideal}}$ and $M_{\text{real}}$ be the corresponding CPTP maps of the ideal and real multiport. For any input state $in$ we have that $M_{\text{ideal}}(in)$ produces a symmetric probability matrix with respect to outcomes of (identical) measurements done by Charlie and Bob. Assume now that Alice wishes Bob to not register a photodetection event while Charlie does so. This probability for the state $M_{\text{ideal}}(in)$ is at most $1/2$. Let the probability of the same event for the state $M_{\text{real}}(in)$ be $d$. In this case, the probability of Alice cheating is $d^{(s_v - s_a)L}$. By the properties of the trace distance (see the "Trace distance and effects (induced probability distributions)" section) we have that $T_D(M_{\text{real}}(in), M_{\text{ideal}}(in)) \geq |1/2 - d|$, where $T_D(\rho, \eta)$ denotes the trace distance between the states $\rho, \eta$. Thus, as long as $D(M_{\text{real}}(in), M_{\text{ideal}}(in)) < 1/2$ the probability of Alice cheating will diminish exponentially quickly in terms of $L$.

One way to conclusively show that $T_D(M_{\text{real}}(in), M_{\text{ideal}}(in)) < 1/2$ holds for our system would be to use full process tomography, which was not performed for our system. (Full process tomography for CV systems is not as well investigated as for qubits, and even qubit process tomography is experimentally demanding.) To evaluate where the actual worst case value $d$ may lie for our implementation we instead analyse how different types of possible imperfections influence this parameter. The imperfections may in general occur within the multiport, but also during the processes leading to Bob and Charlie finally detecting or failing to detect photons, that is, the events of interest (0,1) and (1,0).

In principle $M_{\text{real}}$ can be written as a composition of $M_{\text{ideal}}$ with an additional noise/loss CPTP map collecting all the effects caused by the imperfections in our system. The imperfections characterising the noise/loss map are brought about by the imperfections within the multiport itself. Additionally we also consider the imperfections caused by the realistic authentication/verification process and their effect on the security against repudiation.

If identical sets of equipment are used in Charlie and Bob for the purposes of authentication/verification then the losses and noise induced in the individual arms act as identical and uncorrelated (separable) CPTP maps on the states exiting the multiport. This process can only reduce the trace distance between the reduced states of Bob's and Charlie's systems, thus such noise can only reduce a malevolent Alice's success probabilities.

For the purposes of upper bounding the repudiation probability (i.e. *worst-case scenario*) we may ignore uncorrelated imperfections associated with authentication and verification, and the only imperfections which may help Alice have to lie within the multiport itself. Here, again, any imperfection causing identical uncorrelated noise/loss cannot help Alice, by the arguments above. Hence, we only need to focus on correlated, or differential imperfections inducing correlated or unequal CPTP maps contributing to the cumulative noise/loss map on Bob's and Charlie's reduced states. In our implementation of the multiport, the most likely culprit of differential imperfections comes from the variable air gaps and attenuators placed into the arms of the interferometers. The optical attenuators compensate for different loses in the optical components ensuring the equal intensity of interfering beams. The air gaps compensate for variations in the optical path length in the interferometers which arise from environmental fluctuations. These technical necessities primarily induce an uneven loss in both signal and reference pulses with Bob and Charlie and Bob respectively, and this is the effect we now focus on. This differential loss was studied by the experiment explained in Supplementary Table S1.

We can see in Supplementary Table S1 that differential loss causes Bob to receive on average no less than $1/4$ of the photons compared to Charlie. Since both the signal and the reference pulse are identically attenuated this can, in the worst case scenario, cause the event $(0,1)$ to be ten times more likely than $(1,0)$. If Alice wishes to repudiate her message with the party with the lower loss (Charlie), this induces the worst case value of $d = 4/5$. Even if Bob's and Charlie's output losses were a thousand-fold different (inducing the value $d = 1000/1001$), the forging probability as a function of the signature length L is significantly higher than the refutation probability, which will become clear from the computations to follow. Thus if one is interested in probability of the protocol failing in any way, security against forging (and likewise the required robustness) will constitute the dominant factor in the overall failure probability of the protocol. Forging is therefore the focus of the remainder of this document.

**Security against forgery**

We identify two types of cheating strategies for forger Bob:

- Passive strategy: Bob does not interfere during distribution of the quantum signatures, but tries to cheat by inspecting his copy of the quantum signature. These types of attacks are somewhat analogous to individual and collective attacks in quantum key distribution (QKD).

- Active strategy: Bob is malevolent throughout the distribution of the quantum signatures - this constitutes the most general type of attacks. These attacks are somewhat analogous to coherent attacks in QKD.

We begin with analysis of the passive attack, the results of which will be the crux of the security analysis for active attacks.

**Passive forging - separable attacks**

In this attack, Bob does not interfere throughout the quantum signature distribution phase. To forge a message, he applies one (optimal) measurement to estimate the phase of each of his elements of the quantum signature and sends his best guess to Charlie. To calculate Bob's cheating probability we focus on calculating the probability of Bob not generating a photodetection event with Charlie, per individual quantum signature element. This probability is given by

$$p_{\text{forgery}} = \min_{\{\{\Pi_\phi\}\}} \frac{1}{N} \sum_\phi \sum_\theta Tr(\Pi_\phi \rho^\theta) c_{\phi,\theta} \tag{S10}$$

where
- $Tr(\Pi_\phi \rho^\theta)$ is the probability of Bob measuring (and thus declaring) the angle $\phi$ if the state he measured was encoded with the angle $\theta$.
- $c_{\phi,\theta}$ is the probability of Charlie registering a photodetection event in his signal null-port if the state he had in his memory was encoded with $\theta$ and Bob declared $\phi$.

The expression for $p_{\text{forgery}}$ above is minimised over all possible POVMs. The minimum constitutes the cost of a minimum cost measurement, and the criteria for the minimum are given by[18]

1. $\Gamma = \sum_i \Pi_i W_i = \sum_i W_i \Pi_i$ for $W_i = 1/N \sum_j C_{i,j} \rho_j$ (S11)

2. $\Gamma = \Gamma^\dagger$ (S12)

3. $\Pi_i (W_i - \Gamma) = (W_i - \Gamma) \Pi_i = 0$ for all $i$ (S13)

4. $(W_i - \Gamma)$ is positive-semidefinite for all $i$ (S14)

We refer to the set of conditions above (equations S11 to S14) as Helstrom criteria 1-4, respectively. The cost matrix $C$ with elements $c_{\phi,\theta}$ is obtained from experimental results. For clarity, the cost matrix is indexed according to encoding angles. However, to remain compatible with the indexing tradition for minimum cost measurements, in the abstract formulation of the problem the indexing is performed across integers, so that the index angle $\theta = 2k\pi/N$ corresponds to the integer index $k$.

In the most general case for an arbitrary cost matrix, the computation of the optimal measurement is difficult. However note that if the cost matrix $C$ is replaced by a cost matrix where each entry is less than or equal to the entries of the original cost matrix (an element dominated matrix), the overall cost of the optimal transform can only decrease. In the ideal case, where the experiment is completely symmetric, the cost matrix $C$ is circulant and symmetric. However, in reality it is just close to a symmetric and circulant matrix. If we now substitute the cost matrix with the closest element dominated (element-wise smaller than the cost matrix) symmetric and circulant matrix, and compute the cost for this matrix, by the remark above we have found a lower bound for $p_{\text{forgery}}$. In a similar fashion, we can compute the upper bound for the same expression by considering the symmetric and circulant cost matrix which upper bounds the elements of the actual cost matrix $C$. As we will see, these two values are very close, so the lower bound we will compute is very close to the actual value. This reduction simplifies the computation as for circulant and symmetric positive cost matrices the first, second and third Helstrom criteria are satisfied for the so-called minimum-error or square-root measurement[18]. This statement is proven in the "Statements about minimum cost measurements" section, where the square-root measurement is also given. The conditions on the cost matrix for the fourth Helstrom criterion to be satisfied with the square-root measurement are more involved, and for our experimental results we have verified this criterion numerically.

If Bob were honest, the probability of him triggering a photodetection event with Charlie, per quantum signature element state, in Charlie's signal null-port, is given by the average of the diagonal of the cost matrix $C$. Let this value be $p_{\text{original}}$, and let the corresponding value if Bob is forging be $p_{\text{forgery}}$. We define the gap between these two values as $g = p_{\text{forgery}} - p_{\text{original}}$. If we now set the authentication and verification thresholds at

$$s_a = p_{\text{original}} + 1/3g \qquad (S15)$$

and

$$s_v = p_{\text{forgery}} - 1/3g = p_{\text{original}} + 2/3g \qquad (S16)$$

then the probability of Bob successfully forging the signature is equal to the probability that the fraction of photodetection events is less than $s_v$ where the expected fraction is $p_{\text{forgery}}$. Note that the value $(s_v - s_a)$, appearing in the analysis of security against refutation, equals $1/3g$. $p_{\text{forgery}}$ is then the probability that in a repeated experiment ($M$ times) with a binary outcome with mean $p_{\text{forgery}}$, the normalised measured outcome diverges from the expectancy by more than $p_{\text{forgery}} - s_v = 1/3g$ and this is bounded using Hoeffding's inequalities[24] as follows:

$$\varepsilon_{\text{forging}} = P(\text{Bob cheats}) \leq 2\exp(-\frac{2}{9}g^2 L). \tag{S17}$$

A similar analysis gives us the robustness as well:

$$\varepsilon_{\text{robustness}} = P(\text{Honest setting abort}) \leq \exp(-\frac{2}{9}g^2 L) + \exp(-\frac{4}{9}g^2 L) \tag{S18}$$

which is bounded above by $\varepsilon_{\text{forging}}$.

**Estimation of forging probabilities for the passive attack based on experimental data**

The cost matrix realised by our experimental set-up using 8 differing phase states and with average photon number of $|\alpha|^2 = 0.16$ per pulse is given by

$$C = \begin{pmatrix} 3.89 & 4.40 & 5.24 & 5.95 & 6.35 & 6.00 & 5.29 & 4.39 \\ 4.56 & 3.88 & 4.43 & 5.29 & 6.04 & 6.39 & 6.02 & 5.20 \\ 5.28 & 4.60 & 3.89 & 4.42 & 5.29 & 6.02 & 6.37 & 5.95 \\ 5.68 & 5.22 & 4.58 & 3.90 & 4.40 & 5.24 & 5.91 & 6.30 \\ 6.36 & 5.68 & 5.27 & 4.59 & 3.89 & 4.43 & 5.24 & 6.01 \\ 5.62 & 6.36 & 5.66 & 5.23 & 4.57 & 3.89 & 4.41 & 5.30 \\ 5.26 & 5.68 & 6.40 & 5.70 & 5.22 & 4.60 & 3.88 & 4.40 \\ 4.61 & 5.24 & 5.65 & 6.36 & 5.68 & 5.22 & 4.56 & 3.88 \end{pmatrix} \times 10^{-3}. \tag{S19}$$

The cost matrix is related to the values presented in Figure 4. Whereas in Figure 4, the values presented are the encoding errors, the cost matrix values are calculated by dividing the total number of signal null-port counts by the total number of pulses (including vacuum) emitted by Alice during the duration of the measurement (or equivalently clock frequency multiplied by measurement duration). As in Figure 4, the diagonal elements represent the cases when receiver measures using the same phase as set by Alice, the off-diagonal elements represent the cases where a different phase is employed. The number of pulses reaching a receiver's signal null-port is, roughly speaking, proportional to the intensity of the incident light, and the cost matrix elements will therefore scale linearly with $|\alpha|^2$. The uncertainties in the cost matrix values are element dependent but are of the order of $7 \times 10^{-6}$.

The symmetrised and circularised cost matrix which lower bounds the original cost matrix is characterised by its first row which is given by
$$C'_{\text{row}} = (3.88, 4.39, 5.22, 5.91, 6.30, 5.91, 5.22, 4.39) \times 10^{-3} \tag{S20}$$
and the upper bounding symmetrised and circularised matrix is characterised by the row
$$C''_{\text{row}} = (3.90, 4.43, 5.30, 6.04, 6.39, 6.04, 5.30, 4.43) \times 10^{-3}. \tag{S21}$$

For both lower and upper bounding cost matrices we have numerically checked that the fourth Helstrom criterion is satisfied, so in both cases, the minimum cost measurement is realised by the square-root measurement, and the costs are given by $cost_{lower} = 4.70 \times 10^{-3}$ and $cost_{upper} = 4.76 \times 10^{-3}$. As noted, for the worst case scenario, we need to take the largest diagonal element of the actual cost matrix as $p_{honest}$, which is $3.9 \times 10^{-3}$, and the lower and upper bounds on the gap $g$ are $g_{lower} = 8.03 \times 10^{-4} \pm 0.3 \times 10^{-4}$ and $g_{upper} = 8.64 \times 10^{-4} \pm 0.6 \times 10^{-4}$, This demonstrates that the bounding technique yields a useful bound. Thus the security of our system is characterised by the lower bound on the gap $g_{lower} = 8.03 \times 10^{-4} \pm 0.3 \times 10^{-4}$.

We predicted the cost matrices for a range of different $|\alpha|^2$ values using the theoretical model presented in the Methods section and used these to estimate the gap $g$ for each $|\alpha|^2$ value. A graph of these estimations is presented in Supplementary Figure S1. The values of the diagonal elements of the cost matrix were found to range from a minimum of $3.83 \times 10^{-3} \pm 6 \times 10^{-6}$ (occurring at an $|\alpha|^2$ value of 0.04 and a phase encoding of $\pi/8$) to a maximum of $3.97 \times 10^{-3} \pm 6 \times 10^{-6}$ (occurring at an $|\alpha|^2$ value of 0.28 and a phase encoding of $3\pi/8$). As expected the elements of the cost matrix show a strong linear dependence on the value of $|\alpha|^2$. Calculation of the entries of the cost matrix as the function of the coherent state in the quantum signature and the one which we generated to compare it against is straightforward in the ideal case with perfect detectors. It is given by[16]

$$c_{\phi,\theta}^{ideal} = 1 - \exp\left\{-|\alpha|^2 \sin^2\left[(\phi-\theta)/2\right]\right\} \qquad (S22)$$

**Passive attacks with collective measurements**

In the security analysis for the passive attack above we have assumed that the malevolent Bob performs individual identical measurements on his quantum signature states in order to produce a 'best guess' sequence of phase angles to use when forging a message. A collective measurement may in principle yield a higher probability of forging a message, but here we prove this is not the case. This is not a surprising result as the quantum signature element states are not mutually correlated. Recall, the pivotal value which we used to characterise the security of our system was $p_{forgery}$ - the probability of a cheating Bob not causing a photodetection event during Charlie's verification phase, per individual quantum signature element state. We now show that any average probability of a cheating Bob not causing a photodetection event during Charlie's verification phase, per individual quantum signature state, if Bob uses a global measurement, can be achieved by measurements of individual signature states. This shows that collective measurement strategies cannot help a malevolent Bob.

Let $\{\Pi_{\vec{\phi}}\}, \vec{\phi} = (\phi_1, \ldots, \phi_L)$ be the POVM elements of any global measurement Bob may employ, where the index is a sequence of angles corresponding to Bob's estimate of the angles. Then the average probability of Bob not causing a photodetection event with Charlie is

$$p_{\text{forgery}}^{\text{average}} = \frac{1}{N^L} \sum_{\vec{\phi}} \sum_{\vec{\theta}} Tr(\Pi_{\vec{\phi}} \rho^{\vec{\theta}}) c_{\vec{\phi},\vec{\theta}} \tag{S23}$$

with $\rho^{\vec{\theta}} = \otimes_{k=1}^{L} |e^{i\theta_k}\alpha\rangle\langle e^{i\theta_k}\alpha|$ and $c_{\vec{\phi},\vec{\theta}} = \sum_{k=1}^{L} c_{\phi_k,\theta_k}/L$. Then we have the following derivation:

$$p_{\text{forgery}}^{\text{average}} = \frac{1}{N^L} \sum_{\vec{\phi}} \sum_{\vec{\theta}} Tr(\Pi_{\vec{\phi}} \rho^{\vec{\theta}}) c_{\vec{\phi},\vec{\theta}} =$$

$$\frac{1}{N^L} \frac{1}{L} \sum_{k=1}^{L} \sum_{\phi_k} \sum_{\theta_k} \sum_{(\phi_1,\ldots,\phi_{k-1},\phi_{k+1},\ldots,\phi_L)} \sum_{(\theta_1,\ldots,\theta_{k-1},\theta_{k+1},\ldots,\theta_L)} Tr(\Pi_{\vec{\phi}} \rho^{\vec{\theta}}) c_{\phi_k,\theta_k} =$$

$$\frac{1}{N^L} \frac{1}{L} \sum_{k=1}^{L} \sum_{\phi_k} \sum_{\theta_k} Tr\left( \sum_{(\phi_1,\ldots,\phi_{k-1},\phi_{k+1},\ldots,\phi_L)} \sum_{(\theta_1,\ldots,\theta_{k-1},\theta_{k+1},\ldots,\theta_L)} \Pi_{\vec{\phi}} \rho^{\vec{\theta}} \right) c_{\phi_k,\theta_k} =$$

$$\frac{1}{N} \frac{1}{L} \sum_{k=1}^{L} \sum_{\phi_k} \sum_{\theta_k} Tr\left[ \left( \sum_{(\phi_1,\ldots,\phi_{k-1},\phi_{k+1},\ldots,\phi_L)} \Pi_{\vec{\phi}} \right) \left( \sum_{(\theta_1,\ldots,\theta_{k-1},\theta_{k+1},\ldots,\theta_L)} \frac{1}{N^{L-1}} \rho^{\vec{\theta}} \right) \right] c_{\phi_k,\theta_k}. \tag{S24}$$

Note that the operator

$$\tilde{\Pi}_{\phi_k}^{k} = \left( \sum_{(\phi_1,\ldots,\phi_{k-1},\phi_{k+1},\ldots,\phi_L)} \Pi_{\vec{\phi}} \right) \tag{S25}$$

is a positive operator, and that

$$\left( \sum_{(\theta_1,\ldots,\theta_{k-1},\theta_{k+1},\ldots,\theta_L)} \frac{1}{N^{L-1}} \rho^{\vec{\theta}} \right) = \Phi^{\otimes(k-1)} \otimes \rho^{\theta_k} \otimes \Phi^{\otimes(N-k)} \tag{S26}$$

where $\Phi = 1/N \sum_{\theta} \rho^{\theta}$ is the average quantum signature element state. Thus we have

$$p_{\text{forgery}}^{\text{average}} = \frac{1}{N} \frac{1}{L} \sum_{k=1}^{L} \sum_{\phi_k} \sum_{\theta_k} Tr\left( \tilde{\Pi}_{\phi_k}^{k} \Phi^{\otimes(k-1)} \otimes \rho^{\theta_k} \otimes \Phi^{\otimes(N-k)} \right) c_{\phi_k,\theta_k} =$$

$$\frac{1}{N} \frac{1}{L} \sum_{k=1}^{L} \sum_{\phi_k} \sum_{\theta_k} Tr\left[ \tilde{\Pi}_{\phi_k}^{k} \left( \Phi^{\otimes(k-1)} \otimes \mathbf{1} \otimes \Phi^{\otimes(N-k)} \right) \left( \mathbf{1}^{\otimes(k-1)} \otimes \rho^{\theta_k} \otimes \mathbf{1}^{\otimes(N-k)} \right) \right] c_{\phi_k,\theta_k} \tag{S27}$$

where $\mathbf{1}$ is the identity operator acting on signature element state space. The trace superoperator above can be decomposed into the partial trace over the $k^{\text{th}}$ subsystem and the partial trace over every subsystem except the $k^{\text{th}}$ subsystem, which we will denote $Tr_{\bar{k}}$:

$$p_{\text{forgery}}^{\text{average}} = \frac{1}{N} \frac{1}{L} \sum_{k=1}^{L} \sum_{\phi_k} \sum_{\theta_k} Tr\left[ \tilde{\Pi}_{\phi_k}^{k} \left( \Phi^{\otimes(k-1)} \otimes \mathbf{1} \otimes \Phi^{\otimes(N-k)} \right) \left( \mathbf{1}^{\otimes(k-1)} \otimes \rho^{\theta_k} \otimes \mathbf{1}^{\otimes(N-k)} \right) \right] c_{\phi_k,\theta_k} =$$

$$\frac{1}{N} \frac{1}{L} \sum_{k=1}^{L} \sum_{\phi_k} \sum_{\theta_k} Tr_k \left\{ Tr_{\bar{k}} \left[ \tilde{\Pi}_{\phi_k}^{k} \left( \Phi^{\otimes(k-1)} \otimes \mathbf{1} \otimes \Phi^{\otimes(N-k)} \right) \left( \mathbf{1}^{\otimes(k-1)} \otimes \rho^{\theta_k} \otimes \mathbf{1}^{\otimes(N-k)} \right) \right] \right\} c_{\phi_k,\theta_k} =$$

$$\frac{1}{N} \frac{1}{L} \sum_{k=1}^{L} \sum_{\phi_k} \sum_{\theta_k} Tr_k \left\{ \rho^{\theta_k} Tr_{\bar{k}} \left[ \tilde{\Pi}_{\phi_k}^{k} \left( \Phi^{\otimes(k-1)} \otimes \mathbf{1} \otimes \Phi^{\otimes(N-k)} \right) \right] \right\} c_{\phi_k,\theta_k}. \tag{S28}$$

Since the partial trace is a positive trace preserving superoperator, the operator

$$\Pi_{\phi_k}^{k} = Tr_{\bar{k}} \left[ \tilde{\Pi}_{\phi_k}^{k} \left( \Phi^{\otimes(k-1)} \otimes \mathbf{1} \otimes \Phi^{\otimes(N-k)} \right) \right] \tag{S29}$$

is a positive operator. Moreover, it is easy to verify that $\sum_\phi \Pi_\phi^k = \mathbf{1}$ so the operators $\{\Pi_\phi^k\}_\phi$ comprise a complete set of POVM elements acting on the $k^{\text{th}}$ subsystem. Thus we have

$$p_{\text{forgery}}^{\text{average}} = \frac{1}{L} \sum_{k=1}^{L} \frac{1}{N} \sum_{\phi_k} \sum_{\theta_k} Tr\left(\rho^{\theta_k} \Pi_{\phi_k}^k\right) c_{\phi_k, \theta_k} \qquad (S30)$$

and we have expressed the average probability of a cheating Bob causing a photodetection event with Charlie in terms of individual measurements on quantum signature states, without any assumption on the choice of the global measurement. This concludes our analysis which implies that any cheating probability achieved by a global measurement can be realised by independent single system measurements.

**Security against forging - active attack**

In this section we analyse Bob's forging probabilities in the case he employs an active, separable strategy. In active separable strategies, Bob is allowed to alter the states he sends to Charlie during the quantum signature distribution phase, but his malevolent activity is assumed to be equal for each quantum signature element state, and he also acts individually and identically on each element state. By altering the states he sends to Charlie, Bob can try to increase the probability to successfully forge a message later on. We will call the states Bob sends to Charlie the "response states". Here, to guarantee security, we must take into account Charlie's multiport null-port photodetection events.

For the $k^{\text{th}}$ element of the quantum signature, which has a phase of $\theta_k$, Bob has access to his copy of the quantum signature, along with the "half pulse" he received from Charlie. This can be represented by the state $\left|e^{i\theta_k}\sqrt{3/2}\alpha\right\rangle$ in total. In order to forge a message in the future, Bob will at some stage have to commit to an angle $\phi_k$ which will comprise the forged private key. To select the best angle to commit to for the private key, Bob makes a generalised measurement on a fraction of the state $\left|e^{i\theta_k}\sqrt{3/2}\alpha\right\rangle$, and we allow this fraction to be anything between zero and unity. Without the loss of generality, we may assume that the measurement takes place before Bob sends a response state to Charlie within the multiport, since knowing the result of the measurement can only improve Bob's ability to select a response state that would increase his probability of successfully forging a message. The response state $\eta^{\theta_k, \phi_k}$ may in general depend on both the actual phase value of the $k^{\text{th}}$ quantum signature state and on Bob's measurement outcome. We note that in the case of a passive strategy, the response state will be $\left|e^{i\theta_k}\alpha/\sqrt{2}\right\rangle$. The forwarded, possibly altered response state is then interfered on Charlie's final multiport beamsplitter with Charlie's half of the $k^{\text{th}}$ quantum signature state, and one output arm (the multiport null-port) is measured for a photon count, and the output state of the other arm is stored by Charlie as the $k^{\text{th}}$ quantum signature state. Please see Supplementary Figure S2 for an illustration.

The response state can be written, in the most general $P$-representation form, as $\eta^{\theta_k, \phi_k} = \int P(\beta)|\beta\rangle\langle\beta| d^2\beta$ and the joint state of Charlie's final beamsplitter is then

$$\int_{\mathbb{C}} P(\beta) \underbrace{|\beta/\sqrt{2} - e^{i\theta_k}\alpha/2\rangle\langle\beta/\sqrt{2} - e^{i\theta_k}\alpha/2|}_{\text{null-port}} \otimes \underbrace{|\beta/\sqrt{2} + e^{i\theta_k}\alpha/2\rangle\langle\beta/\sqrt{2} + e^{i\theta_k}\alpha/2|}_{\text{quantum signature element}} d^2\beta . \quad (S31)$$

The probability of detecting no photons at the null-port arm is then

$$Tr(|0\rangle\langle 0| \int_{\mathbb{C}} P(\beta) |\beta/\sqrt{2} - e^{i\theta_k}\alpha/2\rangle\langle\beta/\sqrt{2} - e^{i\theta_k}\alpha/2| d^2\beta) = Tr(|e^{i\theta_k}\alpha/2\rangle\langle e^{i\theta_k}\alpha/2|\eta') \quad (S32)$$

where $\eta' = \int_{\mathbb{C}} P(\beta)|\beta/\sqrt{2}\rangle\langle\beta/\sqrt{2}|d^2\beta$. The state which Charlie will store as the quantum signature element is given by $\hat{D}(e^{i\theta_k}\alpha/2)\eta'\hat{D}^\dagger(e^{i\theta_k}\alpha/2)$, where $\hat{D}(\cdot)$ denotes the displacement operator. The expression on the right hand side of the equality (S32) is sometimes referred to as the expected fidelity between the states $|e^{i\theta_k}\alpha/2\rangle\langle e^{i\theta_k}\alpha/2|$ and $\eta'$. Recall that, in the case of a passive strategy, the state $\eta'$ would be exactly $|e^{i\theta_k}\alpha/2\rangle\langle e^{i\theta_k}\alpha/2|$. From this, it is easy to see that the probability of not detecting a photon at Charlie's multiport null-port is equal to the expected fidelity between the quantum signature elements Charlie will store in the active and passive attack settings, respectively, as illustrated in Supplementary Figure S2.

The process of signature verification for each quantum signature element is a two-outcome measurement the outcomes of which correspond to the detector either registering a photon or not. If a bound of the trace distance between the average stored signature elements in the passive and active attacks can be guaranteed, then we can bound the difference of causing a photon detection event during signature verification for the active and the passive strategies. This is ensured by setting a rejection threshold on the multiport null-port photon event count.

Let $r$ be the fraction of the quantum signature states which have caused a photon event during signature distribution, where the quantum signature is of length $L$. Recall, we are assuming that Bob is acting independently and identically for each signature element state so this fraction can be used to bound the value of $Tr(|e^{i\theta_k}\alpha/2\rangle\langle e^{i\theta_k}\alpha/2|\eta')$ for an average signature element state. Let $x := 1 - Tr(|e^{i\theta_k}\alpha/2\rangle\langle e^{i\theta_k}\alpha/2|\eta')$. Then, by the Hoeffding inequality we have that $P(|x-r| \geq \varepsilon) \leq 2\exp(-2\varepsilon^2 L)$. Thus we have that $1 - Tr(|e^{i\theta_k}\alpha/2\rangle\langle e^{i\theta_k}\alpha/2|\eta') \leq r + \varepsilon$ except with probability $2\exp(-2\varepsilon^2 L)$. The expected fidelity has a well-known relationship with the trace distance[40],

$$T_D(|e^{i\theta_k}\alpha/2\rangle\langle e^{i\theta_k}\alpha/2|, \eta') \leq \sqrt{1 - Tr(|e^{i\theta_k}\alpha/2\rangle\langle e^{i\theta_k}\alpha/2|\eta')} = \sqrt{r + \varepsilon}. \quad (S33)$$

So, we have that the trace distance between the average stored signature elements in the passive and active attacks is less than $\sqrt{r+\varepsilon}$ if the fraction of photodetection events was $r$, except with probability $2\exp(-2\varepsilon^2 L)$. With probability $2\exp(-2\varepsilon^2 L)$ this trace may be above $\sqrt{r+\varepsilon}$ but is always below or equal to unity, due to the properties of the trace distance. Consequently, we can bound the trace distance as follows:

$$T_D(|e^{i\theta_k}\alpha/2\rangle\langle e^{i\theta_k}\alpha/2|, \eta') \leq (1 - 2\exp(-2\varepsilon^2 L))\sqrt{r+\varepsilon} + 2\exp(-2\varepsilon^2 L). \quad (S34)$$

Thus one can ensure that the trace distance between the average stored signature elements in the passive and active attacks is arbitrarily small, by selecting an appropriate *rejection threshold r* and a value $\varepsilon$ for the signature distribution step. Then the trace distance approaches $\sqrt{r+\varepsilon}$ exponentially quickly in the quantum signature length $L$. Let us denote the upper bound on the trace distance by

$$\delta = (1 - 2\exp(-2\varepsilon^2 L))\sqrt{r+\varepsilon} + 2\exp(-2\varepsilon^2 L). \tag{S35}$$

Recall, in the passive attack the probability of not causing a photodetection event per quantum signature state was given by

$$p_{\text{forgery}} = \frac{1}{N}\sum_\phi \sum_\theta Tr(\Pi_\phi \rho^\theta) c_{\phi,\theta}. \tag{S36}$$

For our cost matrix, Bob's optimal measurement was shown to be the square-root measurement, and this was dependant on the structure of the cost matrix $c$. In the active attack, Bob has access to a larger amplitude coherent pulse then in the passive setting, as in principle he can measure Charlie's fraction of the signature states as well. Since we impose a restriction on Bob's activities by checking the multiport null-port count, in practice Bob will not be able to measure out all of the systems he receives, as he is forced to return a perhaps slightly modified variant of half of Alice's signature element, or Charlie's half of the coherent pulse, in order to pass Charlie's null-port test during signature distribution. To lower bound Bob's cheating probabilities we however assume that he can indeed use the entirety of the state he has received from both Alice and Charlie for the measurement. This is equivalent to giving Bob amplified versions of the quantum signatures. We will denote the probability of Bob not causing a photodetection event in a passive strategy with amplified pulses by

$$p_{\text{forgery}}^{\text{amplified, passive}} = \frac{1}{N}\sum_\phi \sum_\theta Tr(\Pi_\phi \rho^\theta_{\text{amplified}}) c_{\phi,\theta} \tag{S37}$$

where

$$\rho^\theta_{\text{amplified}} = \left| e^{i\theta}\sqrt{3/2}\alpha \right\rangle\left\langle e^{i\theta}\sqrt{3/2}\alpha \right|. \tag{S38}$$

The values $c_{\phi,\theta}$ above correspond to the entries of the experimentally obtained cost matrix given explicitly in equation S19.

The induced value $p_{\text{forgery}}^{\text{active}}$ is lower-bounded by the optimisation of the minimum cost problem related to the one above, but where the entries of the cost matrix have been decreased by $\delta$. The following derivation shows that the induced value $p_{\text{forgery}}^{\text{active}}$ deviates from $p_{\text{forgery}}^{\text{amplified,passive}}$ by no more than delta:

$$1/N \sum_\theta \sum_\phi Tr(\Pi_\phi \rho^\theta)([C]_{\phi,\theta} - \delta) =$$
$$1/N \sum_\theta \sum_\phi Tr(\Pi_\phi \rho^\theta)([C]_{\phi,\theta}) - 1/N \sum_\theta \sum_\phi Tr(\Pi_\phi \rho^\theta)(\delta) =$$
$$1/N \sum_\theta \sum_\phi Tr(\Pi_\phi \rho^\theta)([C]_{\phi,\theta}) - 1/N \sum_\theta Tr(\mathbf{1}\rho^\theta)(\delta) =$$
$$1/N \sum_\theta \sum_\phi Tr(\Pi_\phi \rho^\theta)([C]_{\phi,\theta}) - 1/N \sum_\theta (\delta) =$$
$$1/N \sum_\theta \sum_\phi Tr(\Pi_\phi \rho^\theta)([C]_{\phi,\theta}) - \delta. \tag{S39}$$

Thus we have the bound

$$p_{\text{forgery}}^{\text{active}} \geq p_{\text{forgery}}^{\text{amplified, passive}} - \delta. \tag{S40}$$

For illustration purposes, for our experimental set-up, the 'amplified' value is $p_{\text{forgery}}^{\text{amplified, passive}} = 4.61 \times 10^{-3} \pm 7 \times 10^{-6}$, inducing a slightly reduced gap of $g^{\text{amplified}} = 7.13 \times 10^{-3} \pm 3 \times 10^{-5}$ when Bob employs an active forging strategy. Then the overall probability of Bob forging is again given by Hoeffding's inequality as

$$\varepsilon_{\text{forging}} \leq 2\exp(-\frac{2}{9}(g^{\text{amplified}} - \delta)^2 L) \tag{S41}$$

which again approaches zero exponentially quickly in the signature length $L$, as long as we ensure that $\delta < g^{\text{amplified}}$. For robustness when allowing for active attacks we obtain an analogous bound for robustness

$$\varepsilon_{\text{robustness}} \leq \exp(-\frac{2}{9}\left(g^{\text{amplified}} - \delta\right)^2 L) + \exp(-\frac{4}{9}\left(g^{\text{amplified}} - \delta\right)^2 L), \tag{S42}$$

which is less than or equal to $\varepsilon_{\text{forging}}$.

In this analysis we assumed that the detectors at the multiport null-port were perfect. Here we briefly consider the effects of imperfect devices on the security parameters. First, taking into account the known detector losses one can still work out the required rejection threshold $r$ and the required signature length $L$ to ensure $\delta < g^{\text{amplified}}$. The losses will make the required threshold lower, and the accompanying signature length $L$ longer compared to the values obtained in the ideal case previously. Nonetheless, arbitrarily small values of $\delta$ can still be obtained efficiently in the signature length. Additionally, the differential losses occurring within the multiport would also cause Bob and Charlie to have differential sensitivities to cheating. If the polarisation of the light in the polarisation maintaing fibre were to degrade this would result in a decreased interferometric visibility which would have the effect of routing a greater number of pulses to the multiport null-ports (if the degradation occurred in the multiport) and a greater number of pulses to the signal null-ports (if the degradation occurred in a receiver's interferometer). In our experiment, the measurements of the cost matrix given in $C$ in the "Estimation of forging probabilities for the passive attack based on experimental data" section from which the gap $g$ is calculated were performed on the party with the lower overall losses. The party with the losses lowered by a multiplicative constant $c$ would realise a guaranteed value of the gap $g' = cg$. In our case $c > 1/4$, see Supplementary Table S1. Finally, we briefly address the question of the protocol's robustness with respect to the rejection threshold $r$ which limits the acceptable multiport null-port photon counts. In the presence of dark counts this threshold may be breached, even when all parties are honest. In our system the raw dark count probability per emitted pulse per detector stands at approximately $p(\text{dark}) = 3.2 \times 10^{-6}$. To take this into account, the baseline threshold $r$ (chosen to achieve the required levels of security against active strategy forging) should simply be increased by the value of $p(\text{dark})$. The realised security levels are not jeopardised as no cheating response state Bob may send can reduce the dark count rate. The dark count rate is limited by the detector and can only possibly increase from the baseline level realised with a passive strategy. Analogous arguments hold if additional causes for a photon detection event not due to a cheating response state, such as interferometric visibility and background count rate, are considered.

The honest setting rejection probability for such a setting can be again shown to vanish exponentially quickly in terms of the signature length $L$ as

$$\varepsilon_{\text{robustness}}^{\text{multiport}} \leq \exp(-2r^2 L) \,. \tag{S43}$$

**Coherent strategies for active attacks**

Here, we give a plausible argument that coherent, or any type of general strategy Bob may employ does not improve Bob's forging probabilities when compared to the separable cheating strategy, discussed in the previous section. The technique we suggest to show this is analogous, albeit more involved, to the one used in the proof of security against refutation for the coherent attacks. We shall consider two types of fictitious protocols, games, which are obtained by modifying the original protocol, and two types of strategies by Bob: individual and coherent. We will denote the original protocol by "O". A modified original protocol, which we call a "sequential, delayed, with disclosure protocol", we will denote "SDwD". Finally, we will use a modified protocol called a delayed protocol, denoted "D".

In the SDwD protocol, the states sent by Alice are accumulated halfway within the multiport: Bob receives all the original quantum signature element states, along with the 'half' of the pulse from Charlie, and Charlie accumulates all his 'half' pulses. This constitutes the 'delay' in the designation of this modified protocol. From here the protocol continues sequentially: at the $k^{\text{th}}$ step, Bob sends the $k^{\text{th}}$ response state, Charlie interferes it with his corresponding 'half' pulse and obtains the corresponding null-port measurement outcome. At this point, Bob chooses (commits to) a private key element (phase angle) based upon the measurement of whatever system he may have. Without the loss of generality this commitment/measurement could have taken place at the instance of Bob generating the response state. Bob then declares his guess of the private key element, and Charlie proceeds to perform the verification for this signature element. This constitutes the 'sequential' attribute in the protocol designation. Finally, Alice, who is an honest player in this modified game, at this point reveals the actual angle which she encoded in the $k^{\text{th}}$ pulse to the cheater Bob. Note that this happens after the verification for this pulse has been carried out. This procedure is sequentially repeated for all signature elements, and Bob wins the game, an event we shall designate "Bob cheats", if he managed to pass both the null-port and the verification thresholds.

The modified protocol D just introduces the change of first accumulating the states within the multiport (the 'delay' explained above), before continuing, and is otherwise identical to the original protocol O.

Concerning Bob's activities, we distinguish a separable strategy corresponding to individual identical activities, denoted "S", and a coherent strategy "C". In a separable strategy S, Bob chooses a response state, and commits to a to-be-declared private key element ('best guess' phase) by measuring the quantum signature states individually, and an identical strategy is applied for each signature element. Also, the response states are not entangled with each other. The probability $P(\text{Bob cheats} | \text{O,S})$, i.e. the probability of Bob successfully forging using a separable attack in the original protocol, is the value computed in the previous section. In a coherent strategy C, Bob is not restricted in any way, aside from the protocol specifications, which would otherwise cause an implicit protocol abort. An example of this would be Bob's failure to choose a phase angle and declare it to Charlie at any step of the SDwD protocol.

Our goal is to prove the following sequence of (in)equalities:
$$P(\text{Bob cheats} | \text{O,S}) \leq P(\text{Bob cheats} | \text{O,C}) \quad \text{(a)} \tag{S44}$$
$$P(\text{Bob cheats} | \text{O,S}) = P(\text{Bob cheats} | \text{SDwD,S}), \quad \text{(b)} \tag{S45}$$

$$P(\text{Bob cheats} \mid \text{SDwD,S}) = P(\text{Bob cheats} \mid \text{SDwD,C}), \qquad (c) \qquad (S46)$$
$$P(\text{Bob cheats} \mid \text{SDwD,C}) \geq P(\text{Bob cheats} \mid \text{D,C}), \qquad (d) \qquad (S47)$$

and finally

$$P(\text{Bob cheats} \mid \text{D,C}) \geq P(\text{Bob cheats} \mid \text{O,C}). \qquad (e) \qquad (S48)$$

The sequence of inequalities (b) – (e) (equations S45 to S48) then shows $P(\text{Bob cheats} \mid \text{O,S}) \geq P(\text{Bob cheats} \mid \text{O,C})$, which combined with the inequality (a) (equation S44) yields the desired claim,

$$P(\text{Bob cheats} \mid \text{O,S}) = P(\text{Bob cheats} \mid \text{O,C}). \qquad (S49)$$

The first claim (a) (equation S44) is trivial, as a separable strategy is a special case of coherent strategies. The claim (b) (equation S45) is relatively easy as well: due to the separable nature of Bob's attack, neither having all the states at his disposal simultaneously, nor the 'sequential' modification play a role. Since the disclosure of the outcomes and the angles comes after Bob's activity per element state, and since the phases in the quantum signature states are independently and uniformly chosen at random, this information cannot help Bob either. To first sort out the obvious claims, we note that the claim (e) (equation S48) is trivial as well. As Bob can run whatever strategy he would run in the original setting in the delayed setting as well, thus delay can only help, which confirms our claim. For the remainder of this section we will thus be focusing on claims (c) and (d) (equations S46 and S47). We start with the claim (c),

$$P(\text{Bob cheats} \mid \text{SDwD,S}) = P(\text{Bob cheats} \mid \text{SDwD,C}), \qquad (S50)$$

for which we give a plausible motivation, but strictly speaking not a proof.

In the (SDwD, S) setting, Bob accumulates the states he receives within the multiport (from both Alice and Charlie) and then individually acts identically, using a most general physical procedure allowable by quantum mechanics, on each individual state, sequentially producing a response state along with private key element. At each step, this is followed by two measurements by Charlie (null-port and the verification-constituting measurement) and a total disclosure of the originally encoded angle and the measurement outcomes. In contrast, in the (SDwD, C) setting, Bob is allowed to perform any global operation on the states he has, and any ancillary system he may wish to use, but again he has to, at each step, choose a response state and an angle - the private key element. As every response state is processed in run-time, Bob has nothing to gain by using response states which are entangled with his remaining subsystem – since he gets the disclosure information, whatever system he would have following Charlie's measurements, Bob can simply reproduce post disclosure at each step.

We now focus on the first response state and angle Bob will generate, and see whether Bob can increase his probabilities of favourable outcomes of Charlie's measurements by using global operations. Note that the measurement Charlie performs for the first state depends only on the angle encoded within the first quantum signature element sent by Alice. This is independent of all subsequent signature states. Thus no response state, generated based on information Bob may gain by globally measuring the entirety of his system (which includes the entire amplified quantum signature, can influence the outcomes of Charlie's measurements for the first system in a way that benefits Bob, when compared to the separable strategy.

Next, we check whether a coherent strategy starting at the first step can augment his probabilities at later stages. As noted, since all is disclosed post Charlie's measurement at each step, whatever state Bob remains with using the coherent strategy post measurement, Bob can generate using the disclosed information also if using a separable strategy. Thus, at each step, his strategy for that particular step may as well be a separable one, as the full disclosure at each step nullifies any advantage he might have gained by using a coherent strategy. Thus, coherent strategies seem to give Bob no advantage in this setting and claim (c) (equation S46) seems plausible .

To finalise our argument we analyse claim (d) (equation S47) $P(\text{Bob cheats}|\text{SDwD,C}) \geq P(\text{Bob cheats}|\text{D,C})$. The only advantage the protocol variant D may hold for Bob is that he need not commit to a particular private key element angle in run-time, but rather can do a global measurement on his system later. Note that if we assume that there exists no communication between Bob and other parties until Bob wins or loses the game (i.e. cheats or get caught) in the D variant of the protocol then delaying the measurement cannot increase his cheating probability.

In the D protocol Bob does obtain the information whether the multiport null-port threshold has been breached. However, since both the verification and the multiport null-port thresholds violations constitute Bob losing, his measurement strategy should not be conditional on whether he passed the multiport threshold.

Assuming this holds, then in the D protocol, we may assume that Bob measures his system at any point up to the moment when he sends away his last response state. Any measurement Bob may perform in the D protocol can be realised by a large unitary acting on the entirety of his system and a sufficient amount of ancillary systems, followed by single system measurements. This can be seen as a consequence of the Naimark dilation theorem[41]. The single system measurement outcomes will give Bob's choices of private key elements. However, Bob, if he employs a coherent strategy, may perform the same map in the SDwD setting as well, and measure the angle-carrying subsystems sequentially as he is required by the protocol. Bob can thus obtain the same cheating success probability in the SDwD setting as in the D setting by simply ignoring the information he gets from the disclosure in the SDwD setting. But then clearly, gaining additional information can only help Bob, and we have our inequality (d) (equation S47). More formal variants of the proof of claims (c) and (d) (equations S46 and S47) we leave for further research. This concludes our argument.

## Supplementary Methods

In this section we consider the lemmas and related statements required to analyze the security of the system.

### Hoeffding's inequalities

Here we briefly state Hoeffding's inequalities[24] and explain how they are used in the security analysis. We are presenting a special version of these inequalities, which is more directly applicable to our setting.

**Lemma 1.** *Let $X_1, \ldots, X_L$ be independent random variables each attaining values 0 or 1. Let $\bar{X} = 1/L \sum X_i$ be the empirical mean of the variables, and let $E(X)$ be the expectancy of the empirical mean. Then we have*

$$P(\bar{X} - E(\bar{X}) \geq t) \leq \exp(-2t^2 L) \quad \text{(S51)}$$

$$P(|\bar{X} - E(\bar{X})| \geq t) \leq 2\exp(-2t^2 L). \quad \text{(S52)}$$

In the case when we analyse Bob's forgery probabilities, we compute the minimal probability of obtaining a photodetection event on the multiport $p_{cheat}$ which defines a sequence of $L$ random variables as in the statement of the theorem above. Then we set a threshold at $s_v L$, and calculate the probability of an empirical mean of the random variables above diverging from the expectancy by more than $p_{cheat} - s_v$, as this is the requirement for the forgery to be accepted. This corresponds to the second inequality (as the empirical value needs to be below the mean/expectancy. Robustness is calculated similarly, however we need to take into account that both Bob and Charlie could reject if all parties are honest. So to upper bound the probability of abort in the honest setting (using the union bound) we add the two probabilities. In the case of coherent cheating we can compute the average probability of getting a photodetection event, and use the Hoeffding theorem for the induced 'averaged' random variables.

### Trace distance and effects (induced probability distributions)

For the trace distance between two states $T_D(\sigma, \rho)$ we have the property

$$T_D(\sigma, \rho) \geq \frac{1}{2} \sum_x |Tr(\Pi_x \rho) - Tr(\Pi_x \sigma)| \quad \text{(S53)}$$

for any set of POVM elements $\{\Pi_x\}_x$. In our case $\rho$ is the perfectly symmetric subsystem, $\sigma$ represents the physical state attainable in the lab, and the POVM is the four-outcome POVM giving the possible outcomes of photodetection on their individual subsystems. Assume that a malevolent Alice's target result is that Bob accepts and Charlie rejects, hence she wishes to maximise the probability of the outcome $(0,1)$. Let $p_x := Tr(\Pi_x \rho)$, and $q_x := Tr(\Pi_x \sigma)$ for $x \in \{(0,0), \ldots, (1,1)\}$. So we have

$$T_D(\sigma, \rho) \geq \frac{1}{2} \sum_x |p_x - q_x| = \frac{1}{2} |p_{(0,1)} - q_{(0,1)}| + \frac{1}{2} \sum_{\text{all but } (0,1)} |p_x - q_x|. \quad \text{(S54)}$$

Note that if $|p_{(0,1)} - q_{(0,1)}| = \epsilon$ then $\sum_{\text{all but }(0,1)} |p_x - q_x| \geq \epsilon$, leading to

$$T_D(\sigma,\rho) \geq |p_{(0,1)} - q_{(0,1)}|. \tag{S55}$$

From this we have the claim referred to in the section "Security against repudiation with realistic devices".

**Statements about minimum cost measurements**

Here we prove the technical statements from the "Passive forging - separable attacks" section. First we standardise the notation.

- Received states: These are the states Bob receives from Alice and Charlie jointly, so if the amplitude of the individual states of the unperturbed states is $\alpha$, Bob will in total receive the states $|v_k\rangle = |e^{(2k\pi I/N)}\sqrt{3/2}\alpha\rangle$.

- Standard basis: For the states $|v_k\rangle = |e^{(2k\pi I/N)}\sqrt{3/2}\alpha\rangle$ one can show that the states $|b_k\rangle := 1/\sqrt{\lambda_k N}\sum_{l=0}^{N-1}\exp(-2kl\pi I/N)|v_l\rangle$ form an orthonormal basis, where $\lambda_k$ are the eigenvalues of the Gram matrix of the states $|v_k\rangle$. The values $\lambda_k$ are also the diagonal elements (of the diagonal matrix) representing the operator $\sum_{k=0}^{N-1}|v_k\rangle\langle v_k|$ in the orthonormal basis above. The symbol $I$ here denotes the imaginary unit. In this basis the states $|v_k\rangle$ have the f expansion

$$|v_k\rangle := 1/\sqrt{N}\sum_{l=0}^{N-1}\exp(2kl\pi I/N)\sqrt{\lambda_l}|b_l\rangle$$ and, in particular, in this basis all entries for the vector $|v_0\rangle$ are positive.

- The unitary characterising the symmetry of the system is $U$, such that $|v_k\rangle = U^k|v_0\rangle$. In the standard basis this unitary is diagonal: $U = \sum_{l=0}^{N-1}\exp(2\pi l I/N)|b_k\rangle\langle b_k|$.

- With DFT we denote the discrete Fourier transform matrix of (implicit) size $N$, defined element-wise by $[DFT]_{p,q} = \exp(-2\pi I pq/N)$ for $p = 0...N-1$, $q = 0,...,N-1$.

**Lemma 2** *If the input states are symmetric, and the cost matrix is circulant, then Helstrom condition 3. holds for the square-root measurement.*

**Proof:** We have the risk operators defined as

$$W_i = \frac{1}{N}\sum_j c_{i,j}|v_j\rangle\langle v_j| = \frac{1}{N}\sum_j c_{i,j}U^j|v_0\rangle\langle v_0|U^{-j}. \tag{S56}$$

If the cost matrix $C = [c_{i,j}]_{i,j}$ is circulant we have that

$$U^k W_0 U^{-k} = W_k. \tag{S57}$$

The Lagrangian operator is defined as

S25

$$\Gamma = \sum_i \Pi_i W_i \tag{S58}$$

The square-root measurement is defined by the operators

$$\Pi_i = \Phi^{-1/2} |v_i\rangle\langle v_i| \Phi^{-1/2} \tag{S59}$$

where

$$\Phi = \sum_i |v_i\rangle\langle v_i| = \sum_i U^i |v_0\rangle\langle v_0| U^{-i}. \tag{S60}$$

We will often use the following property:

**Lemma 2.a** *For any square matrix $A$ we have that*

$$\sum_i U^i A U^{-i} = N A' \tag{S61}$$

*where $A'$ is the diagonal matrix containing the main diagonal of $A$, and $N$ is the size of the matrices $U$ and $A$.*

**Proof:** Let $|\omega_l\rangle = \sum_k \exp(2\pi I k l / N) |b_k\rangle$. The ket $|\omega_l\rangle$ in the standard basis contains the main diagonal of the matrix $U^l$. Then it is easy to see that for all square matrices $A$ we have that $U^l A U^{-l} = A \circ |\omega_l\rangle\langle\omega_l|$, where $\circ$ denotes the Hadamard (Shur, point-wise) matrix product[42], which is distributive with respect to matrix addition. Then we have

$$\sum_i U^i A U^{-i} = \sum_i A \circ |\omega_i\rangle\langle\omega_i| = A \circ \sum_i |\omega_i\rangle\langle\omega_i|. \tag{S62}$$

Using the properties of the sums of roots of unity we have that $\sum_i |\omega_i\rangle\langle\omega_i| = N\mathbf{1}$ where $\mathbf{1}$ is the identity matrix. Hence we have proven Lemma 3, as Hadamard-multiplying any matrix with the identity simply eliminates all off-diagonal elements. □

So we have that $\Phi = \sum_i U^i |v_0\rangle\langle v_0| U^{-i} = N |v_0\rangle\langle v_0| \circ \mathbf{1}$. Thus $\Phi$ is diagonal, and by the form of the ket $|v_0\rangle$ it simply collects the eigenvalues of the Gram matrix of the input states across the diagonal. But then $\Phi^{-1/2}$ contains the inverses of the roots of the eigenvalues $\lambda_k$ across the diagonal. Since $U$ is also diagonal, $U$ and $\Phi$ and $\Phi^{-1/2}$ commute, so we have that $\Pi_k = U^k \Pi_0 U^{-k}$, and for the Lagrangian we have that $U^k \Gamma U^{-k} = \Gamma$.

We will also use a slightly more involved lemma, which generalises lemma 2.a:

**Lemma 2.b.** *For any square matrix $A$, and a sequence of $N$ complex numbers $(c_i)_{i=0}^{N-1}$ we have that*

$$\sum_i c_i U^i A U^{-i} = A \circ B \tag{S63}$$

*where $B$ is a circulant matrix, and its first row is given with $DFT.[c_0,\ldots,c_{N-1}]^T$, i.e. the discrete Fourier transform of the vector with entries $c_i$.*

**Proof:** Similar to the proof of the simpler lemma 2.a, with realisation that $\sum_i c_i |\omega_i\rangle\langle\omega_i|$ is a circulant matrix, and its first row is given with $DFT \cdot [c_0, \ldots, c_{N-1}]^T$.

Lemma 2.b is applied to the risk operator $W_0$ to obtain that
$$W_0 = |v_0\rangle\langle v_0| \circ B \tag{S64}$$
where $B$ is a circulant matrix where the first row comprises the eigenvalues of the cost matrix. To see this simply note that the cost matrix is circulant, and the eigenvalues of a circulant matrix are given by the DFT of the first row of the matrix. We need to show that $\underbrace{(W_i - \Gamma)\Pi_i = 0}_{eq.1} = \underbrace{0 = \Pi_i(W_i - \Gamma)}_{eq.2}$. Because of the symmetries we have that $(W_i - \Gamma)\Pi_i = 0$ if and only if $(W_0 - \Gamma)\Pi_0 = 0$ and the analogous holds for the second equality above.

To prove lemma 2 one shows that the following equalities hold
$$W_0 \Pi_0 = \Gamma \Pi_0 \tag{S65}$$
$$\Pi_0 W_0 = \Pi_0 \Gamma. \tag{S66}$$
using the lemmas 2.a, 2.b and the properties listed at the beginning of this section. We omit this derivation as it is a simple yet space consuming.

**Lemma 3.** *If the cost matrix is positive, symmetric and circulant then the first and second Helstrom criteria are satisfied for the minimum error (square-root) measurement for our problem.*

**Proof:** Since $W_i$ is a sum of positive operators with positive weights it is positive, and in particular Hermitian. The operators $\Pi_i$ are positive as they are POVM elements, hence positive and Hermitian as well. Thus we have
$$\Gamma = (\Gamma^\dagger)^\dagger = ((\sum_i \Pi_i W_i)^\dagger)^\dagger = (\sum_i W_i^\dagger \Pi_i^\dagger)^\dagger = (\sum_i W_i \Pi_i)^\dagger \tag{S67}$$
and so the first and second Helstrom conditions given in the "Passive forging - separable attacks" section are equivalent. Thus it suffices to show that
$$\sum_i W_i \Pi_i = \sum_i \Pi_i W_i. \tag{S68}$$
We have that
$$\sum_i W_i \Pi_i = \sum_i U^i W_0 \Pi_0 U^{-i} = N(W_0 \Pi_0) \circ \mathbf{1} \tag{S69}$$
$$\sum_i \Pi_i W_i = \sum_i U^i \Pi_0 W_0 U^{-i} = N(\Pi_0 W_0) \circ \mathbf{1}. \tag{S70}$$
So lemma 3 holds if and only if $(W_0 \Pi_0)$ and $(\Pi_0 W_0)$ have equal diagonal elements. We have shown that
$$W_0 = |v_0\rangle\langle v_0| \circ B \tag{S71}$$
where $B$ is a circulant matrix where the first row comprises the eigenvalues of the cost matrix. Note that for the square root measurement we have the following property:
$$\Phi^{-1/2} |v_0\rangle\langle v_0| \Phi^{-1/2} |b_k\rangle = 1/\sqrt{N} \sum_l |b_l\rangle. \tag{S72}$$

Thus we have for the $k^{th}$ diagonal element of $(W_0 \Pi_0)$ that

$$1/\sqrt{N} \langle b_k | (|v_0\rangle\langle v_0| \circ B)(\sum_l |b_l\rangle) \tag{S73}$$

which is the sum of the elements of the representation of the bra $\langle b_k | (|v_0\rangle\langle v_0| \circ B)$ in the standard basis scaled by $1/\sqrt{N}$. For the $k^{th}$ diagonal element of $(\Pi_0 W_0)$ we get

$$1/\sqrt{N} \sum_l \langle b_l | (|v_0\rangle\langle v_0| \circ B) | b_k \rangle. \tag{S74}$$

Since the matrix $|v_0\rangle\langle v_0|$ is real in the standard basis, these two expressions are equal if $B$ is real and symmetric. Recall, the matrix $B$ is the circulant matrix comprising the eigenvalues of the cost matrix. These are real if and only if the cost matrix is symmetric which we have by the assumption of the lemma. The symmetricity of $B$ is a consequence of the cost matrix comprising real elements. Thus lemma 3 holds.

## Supplementary References